\documentclass[aps,prl,twocolumn,preprintnumbers,superscriptaddress,dblfloatfix,nofootinbib]{revtex4-1}
\usepackage[utf8]{inputenc}
\usepackage{lmodern}
\usepackage{amsmath,amssymb,mhchem}
\usepackage{graphicx}
\usepackage{url}
\usepackage{color}
\usepackage{enumitem}
\usepackage{subfigure}
\usepackage[dvipsnames]{xcolor}
\usepackage[colorlinks=true,breaklinks=true]{hyperref}
\hypersetup{allcolors=[rgb]{0.0 0.0 0.6},linkcolor=[rgb]{0.75 0.05 0.05}}
\usepackage{bm}
\usepackage{epsfig}
\usepackage{amsmath}
\usepackage{amssymb}
\usepackage{slashed}
\usepackage{color}
\usepackage{accents}
\usepackage[dvipsnames]{xcolor}
\usepackage[colorlinks=true,breaklinks=true]{hyperref}
\hypersetup{allcolors=[rgb]{0.0 0.0 0.6},linkcolor=[rgb]{0.75 0.05 0.05}}

\begin{document}

\preprint{IPMU22-0064, KEK-QUP-2022-0017, KEK-TH-2476, KEK-Cosmo-0303}

\title{Signatures of a High Temperature QCD Transition in the Early Universe
}

\author{Philip Lu}
\affiliation{Center for Theoretical Physics, Department of Physics and Astronomy, Seoul National University, Seoul 08826, Korea}

\author{Volodymyr Takhistov}
\affiliation{International Center for Quantum-field Measurement Systems for Studies of the Universe and Particles (QUP, WPI),
High Energy Accelerator Research Organization (KEK), Oho 1-1, Tsukuba, Ibaraki 305-0801, Japan}
\affiliation{Theory Center, Institute of Particle and Nuclear Studies (IPNS), High Energy Accelerator Research Organization (KEK), Tsukuba 305-0801, Japan
}
\affiliation{Kavli Institute for the Physics and Mathematics of the Universe (WPI), UTIAS, \\The University of Tokyo, Kashiwa, Chiba 277-8583, Japan}

\author{George M. Fuller}
\affiliation{Department of Physics, University of California, San Diego, La Jolla, CA 92093}

\date{\today}

\begin{abstract}
Beyond Standard Model  extensions of QCD could result in quark and gluon confinement occurring well above a temperature of $\sim$GeV. These models can also alter the order of the QCD phase transition. The enhanced production of primordial black holes (PBHs) that can accompany the change in relativistic degrees of freedom at the QCD transition therefore could favor the production of PBHs with mass scales smaller than the Standard Model QCD horizon scale. Consequently, and unlike PBHs associated with a standard GeV-scale QCD transition, such PBHs can account for all the dark matter abundance in the unconstrained asteroid-mass window. This links beyond Standard Model modifications of QCD physics over a broad range of unexplored temperature regimes ($\sim 10-10^3$~TeV) with microlensing surveys searching for PBHs. Additionally, we discuss implications of these models for gravitational wave experiments.
We show that a first order QCD phase transition at $\sim7$ TeV is consistent with the Subaru Hyper-Suprime Cam candidate event, while a $\sim 70$ GeV transition is consistent with OGLE candidate events, and also could account for the claimed NANOGrav gravitational wave signal.
\end{abstract}

\maketitle


With tremendous predictive success and extensive experimental testing, the Standard Model (SM) is a central pillar of modern science. 
The interaction of quarks and gluons as described by SM quantum chromodynamics (QCD), predicts that those particles will be in a quark-gluon plasma at early universe temperatures sufficiently above the QCD energy scale $\Lambda_{\rm QCD} \sim 160$~MeV. As the universe expands and the temperature drops below $\Lambda_{\rm QCD}$, chiral symmetry will be broken and the quarks and gluons will be confined in color singlets (e.g., mesons, nucleons, etc.). This is the QCD transition. 
Lattice calculations that employ the high entropy believed to characterize the early universe show that the SM QCD transition is simply a cross-over and not a first order phase transition~\cite{Bhattacharya:2014ara}. However, despite significant progress, lattice methods remain limited and cover only part of the parameter space of the QCD phase diagram.

Astrophysical environments can offer probes of QCD that are complementary to laboratory experiments like relativistic heavy ion collisions.
For example, neutron star mergers provide tests of strong interaction physics at extreme densities~\cite{Annala:2019puf} and low entropy-per-baryon. 
The early universe with its high temperatures and low net baryon density (high entropy-per-baryon $s \sim {10}^{10}$ Boltzmann's constant per baryon) provides a potentially unique probe of QCD. This regime of temperature and entropy is unexplored in laboratory experiments. 
Current observations really only constrain the physics of the early universe that affects neutrino decoupling and primordial nucleosynthesis. These occur at temperature scale  $T \lesssim5$~MeV~\cite{Kawasaki:2000en,deSalas:2015glj,Hasegawa:2019jsa}. The cosmological QCD phase transition has been shown to affect primordial nucleosynthesis~\cite{Fuller:1987ue} but these effects are most important when the transition is first order and the distribution of entropy is inhomogeneous. However, new physics beyond the SM could significantly alter the history of the early Universe.

A QCD phase transition in the early Universe occuring in the SM at $T \sim 160$~MeV significantly affects the equation of state (EoS) governing behavior of the cosmological fluid. The dimensionless parameter $w = p/\rho$, where $p$ is the pressure and $\rho$ is the density, is significantly decreased compared to radiation-dominated environments with $w = 1/3$. Since the pressure balancing gravity becomes weaker at QCD transition, inhomogeneities associated with cosmological perturbations re-entering the horizon collapse with enhanced rate and can naturally lead to primordial black holes with masses peaked in the range associated with these scales~\cite{Byrnes:2018clq,Jedamzik:1998hc,Jedamzik:2020omx,Carr:2019kxo,Carr:2019hud,Dolgov:2020sov,Franciolini:2022tfm,Juan:2022mir}. Interestingly, since the horizon mass at the transition temperature is $\mathcal{O}(1)M_{\odot}$, the resulting solar-mass PBHs have been associated with some of the recent gravitational wave (GW) events of LIGO-Virgo-KAGRA (LVK)~(e.g.~\cite{Clesse:2020ghq}). Stellar-mass PBHs formed prior to galaxies and stars~(e.g.~\cite{Zeldovich:1967,Hawking:1971ei,Carr:1974nx,GarciaBellido:1996qt,Green:2000he,Khlopov:2008qy,Frampton:2010sw,Cotner:2019ykd,Cotner:2018vug,Green:2016xgy,Kusenko:2020pcg,Sasaki:2018dmp,Carr:2020gox,Green:2020jor,Escriva:2022duf}) have been directly linked with LVK GW events more generally~(e.g.~\cite{Nakamura:1997sm,Clesse:2015wea,Bird:2016dcv,Raidal:2017mfl,Eroshenko:2016hmn,Sasaki:2016jop,Clesse:2016ajp,Fuller:2017uyd,Takhistov:2017nmt,Takhistov:2017bpt,Franciolini:2021tla}). In the mass-range associated with the SM QCD transition, PBHs are already constrained from a variety of observations and cannot constitute all of the DM abundance (e.g.~\cite{Sasaki:2018dmp,Carr:2020gox,Green:2020jor}).

In this work, we show that by utilizing PBHs as proxies, a variety of telescope surveys can probe the cosmological QCD transition over unexplored regimes covering orders of magnitude in temperature higher than that of the SM $\Lambda_{\rm QCD} \sim 160$~MeV. Unlike the SM, such a QCD transition could be first order, which would further enhance PBH production due to reduced sound speed and pressure of the cosmological fluid. First order QCD transition is a general expectation from effective field theory if the number of light quarks $N_f \geq 3$~\cite{Pisarski:1983ms}. For example, an ultra-light scalar field and additional massless quarks at the QCD transition allow for a first order transition at lower temperatures $\Lambda_{\rm QCD}$, below $\sim100$~MeV, resulting in PBHs in the LVK mass-window~\cite{Davoudiasl:2019ugw}. We note that while the PBH formation we discuss can also be associated with hidden dark sector gauge dynamics (e.g.~\cite{Gross:2021qgx}), it is particularly intriguing to explore formation within the context of a high temperature QCD transition.
This is because in a significant change in 
 relativistic degrees of freedom in a QCD transition. Additionally, there remains a large unexplored parameter space of one of the most fundamental forces we know.

A high temperature first order QCD transition can readily appear in classes of models where strong coupling becomes a dynamical quantity \cite{Ipek:2018lhm,Croon:2019ugf,Berger:2020maa}. Here, QCD confined at a high temperature scale can dynamically transition at lower temperature to SM QCD with
the standard parameters.
A minimal realization of this is based on a SM gauge singlet scalar $S$ coupling to the gluon field with strength $G^{\mu\nu}$. This could appear in scenarios with radion or dilaton fields, in models based on extra dimensions (e.g.~\cite{vonHarling:2017yew}), or just with an extra scalar coupled to gluons via vector-like fields charged under $SU(3)$. Here, we consider the SM gluon kinetic term
\begin{equation}
\label{eq:modqcd}
    \mathcal{L} \supset -\dfrac{1}{4}\Big(\dfrac{1}{g_{s0}^2} + \dfrac{S}{M}\Big)G_{\mu\nu}^{a}G_a^{\mu\nu} + \dots~,
\end{equation} 
where $g_{s0}$ is the $SU(3)$ gauge coupling when $\langle S \rangle = 0$, and $M$ is the scale characterizing non-renormalizable scalar-gluon interaction.
 
In this scenario, when $S$ acquires a vacuum expectation value (VEV) $\langle S \rangle \neq 0$, the effective modified strong coupling in Eq.~\eqref{eq:modqcd} can be realized. Following renormalization running of the coupling $g_{s0}$ at one loop and considering that QCD confinement occurs when strong the strong coupling constant is $\alpha_s^{-1} \sim 0$, the confinement scale $\Lambda$ is given by~\cite{Ipek:2018lhm}
\begin{equation}
\Lambda(\langle S \rangle) = \Lambda_0 \textrm{Exp}\Big[ \dfrac{24 \pi^2}{2 N_f - 33} \dfrac{\langle S \rangle}{M}\Big]~.
\end{equation}
Taking $N_f = 6$ massless quarks and energy scale $\Lambda_0\sim$GeV that determines $g_{s0}$ 
in the ultraviolet limit, $\langle S \rangle/M = -0.81$ gives a resulting QCD confinement scale of $\Lambda \sim 10$~TeV. A full model potential $V(S)$, which in generality could include at zero temperature distinct $S^n$ terms (with $n = 1,2,3,4$), could readily yield the desired $\langle S \rangle$ for a choice of couplings. Further, we assume that the possible mixing terms between $S$ and the Higgs are small, consistent with Large Hadron Collider results~\cite{ATLAS:2019nkf}.

 \begin{figure*}[t]
\begin{center}
\includegraphics[width=1\columnwidth]{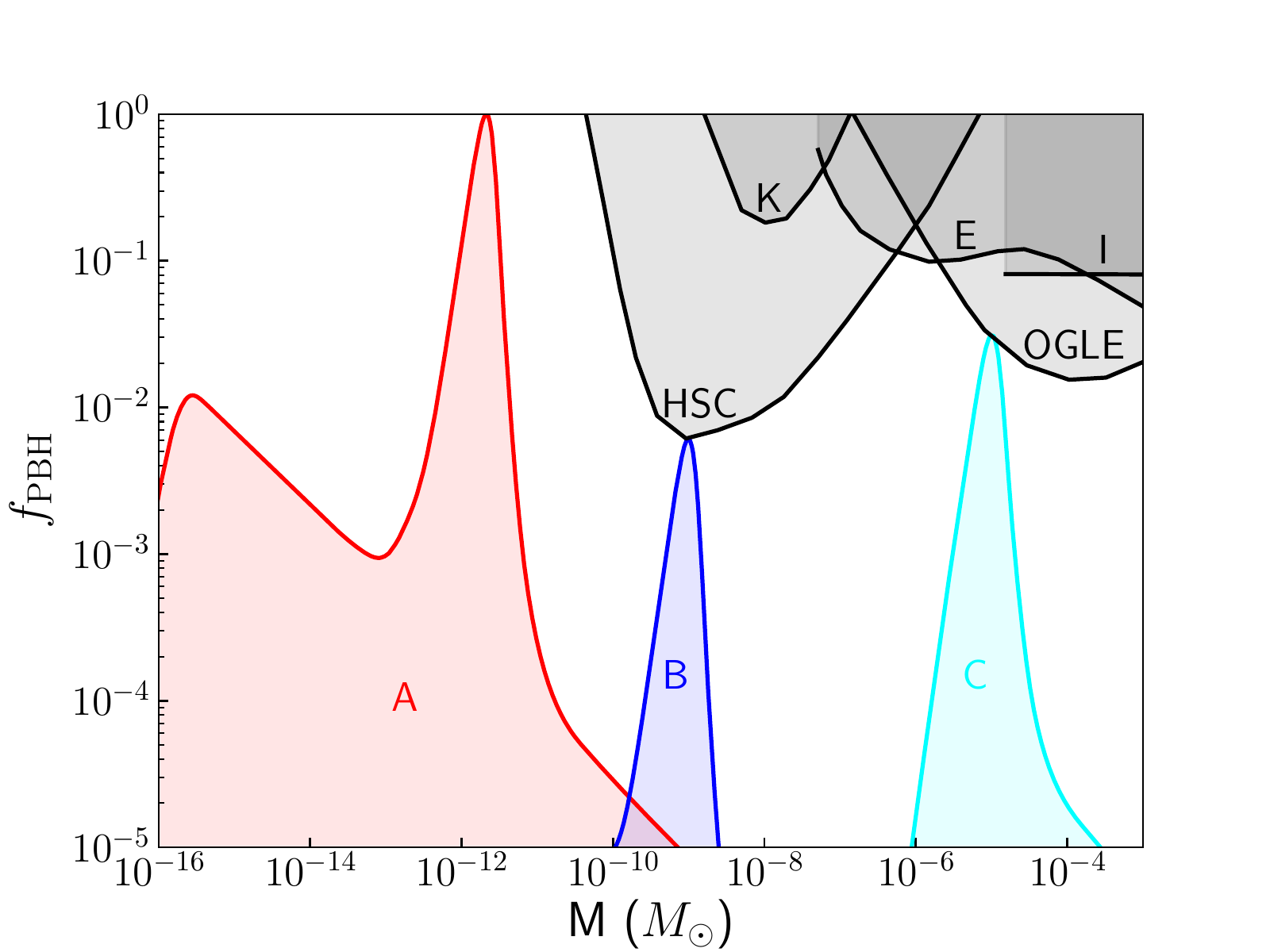} 
\includegraphics[width=1\columnwidth]{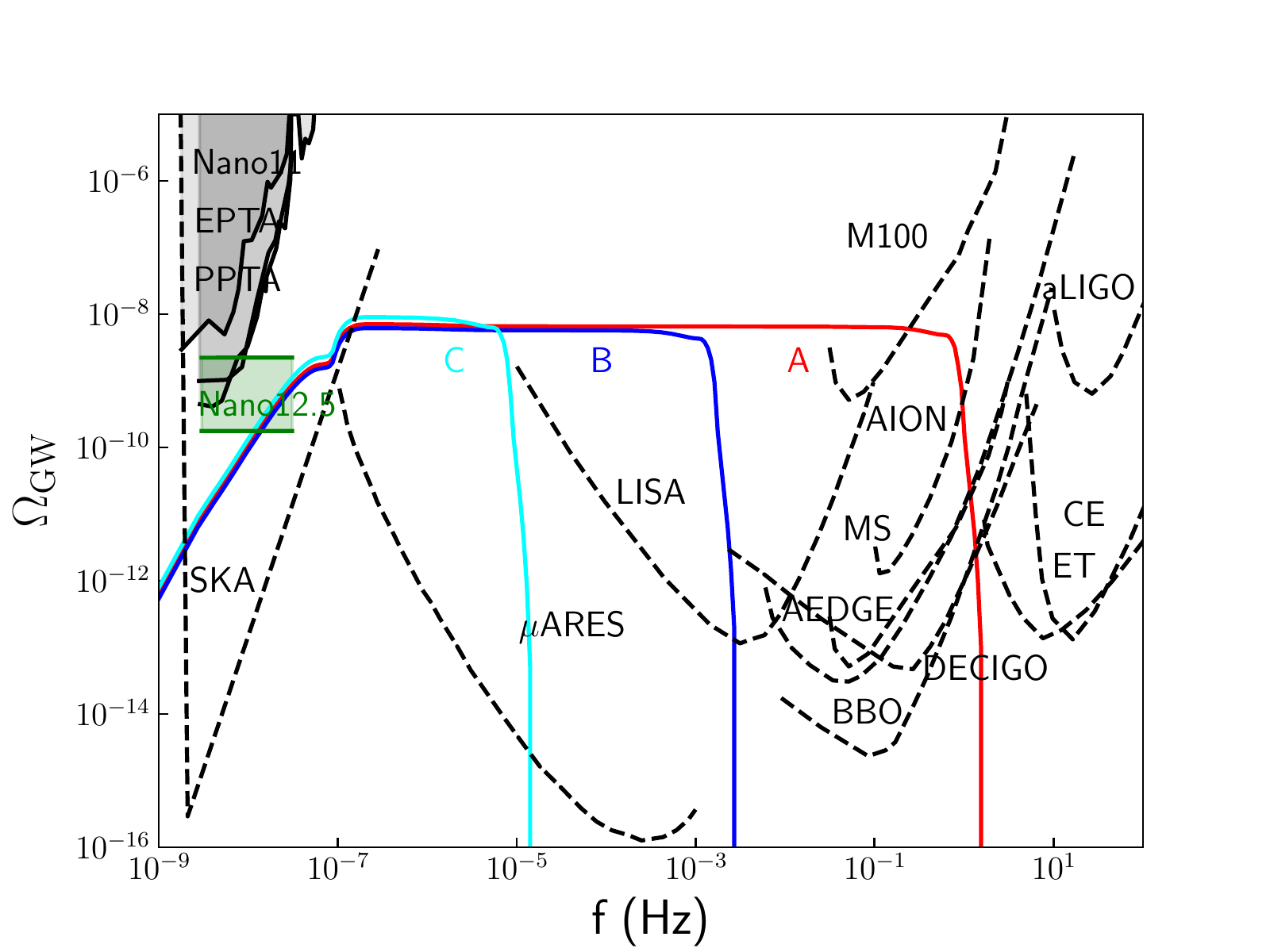}
\caption{
[Left] We show PBH fractional contribution to DM, $f_{\rm PBH}$, for PBHs of mass $M$ in solar masses, in models of a first order QCD transition at high temperature scales. Shown are existing PBH parameter constraints from Subaru Hyper-Suprime Cam (HSC)~\cite{Niikura:2017zjd,Smyth:2019whb}, MACHO/EROS (E)~\cite{Macho:2000nvd,MACHO:2000bzs}, OGLE~\cite{Niikura:2019kqi}, Icarus (I) ~\cite{Oguri:2017ock}, Kepler data (K)~\cite{Griest:2013aaa}. Models A, B and C correspond to those in Table I. [Right] The induced GW spectrum is shown as a function of GW frequency $f$ along with current constraints of EPTA~\cite{Lentati:2015qwp}, PPTA~\cite{Shannon:2015ect}, NANOGrav 11 yrs~\cite{NANOGRAV:2018hou,Aggarwal:2018mgp} as well as projections for SKA~\cite{Zhao:2013bba}, LISA~\cite{LISA:2017}, DECIGO/BBO~\cite{Yagi:2011wg}, Cosmic Explorer (CE)~\cite{LIGOScientific:2016wof}, Einstein Telescope (ET)~\cite{Moore:2014lga,Sathyaprakash:2009xs}, $\mu$-Ares~\cite{Sesana:2019vho}, Magis-100 (M100)/Magis-Space (MS)~\cite{Coleman:2018ozp}, AEDGE~\cite{AEDGE:2019nxb}, AION~\cite{Badurina:2019hst}, aLIGO~\cite{LIGOScientific:2016jlg}.  
}
\label{fig:PBHfull}
\end{center} 
\end{figure*} 

  \begin{table*}[t]  
\centering 
\begin{tabular}{|c|c|c|c|c|c|c|c|c|}  
\hline 
Model & $T_0$ & $\Lambda_{\rm p}$ & $N_{f}$ & $A_s$ & $k_{\rm max}$ & $k_{\rm min}$ & $f_{\rm PBH}$ & $M_{\rm PBH, peak}$ \\
\hline
A & 150 TeV & 450 TeV & 6 & 0.0153 & $5.6\times10^{14}\textrm{ Mpc}^{-1}$ & $5\times10^{7}\textrm{ Mpc}^{-1}$ & 1 & $10^{-12} M_\odot$  \\
\hline
B & 7 TeV & 21 TeV & 6 &   0.0143 &  $10^{12}\textrm{ Mpc}^{-1}$ & $5\times10^{7}\textrm{ Mpc}^{-1}$ & $5*10^{-3}$  & $10^{-9} M_\odot$ \\
\hline
C & 70 GeV & 210 GeV & 5 & 0.0173 &  $5 \times 10^{9}\textrm{ Mpc}^{-1}$ & $5\times10^{7}\textrm{ Mpc}^{-1}$ & $2*10^{-2}$  & $10^{-5} M_\odot$ \\
\hline
\end{tabular}
\caption{Input parameters for QCD transition Models A and B. 
Parameters are
PNJL transition critical temperature $T_0$, momentum cutoff $\Lambda_{\rm p}$, number of massless quarks $N_f$ as well as curvature power spectrum amplitude $A_s$, cut off scales $k_{\rm max}$ and $k_{\rm min}$ that describe the range of masses of resulting PBH distribution associated with corresponding horizon mass.}
\end{table*}

The implication of a QCD phase transition at extreme temperatures, above the electroweak scale, is largely unexplored. To gain insight, we first model it using effective Polyakov-loop enhanced Nambu–Jona-Lasinio (PNJL) theory~\cite{Fukushima:2003fw} following Ref.~\cite{Ratti:2005jh,Helmboldt:2019pan}. This treatment can model the SM QCD transition~\cite{Ratti:2005jh}. We reproduce the PNJL results for SM QCD as confirmed by lattice calculations. We then calculate the equation of state (EoS) behavior when the critical temperature $T_c$ associated with the phase transition is set to 1 TeV. This scale is beyond that of electroweak phase transition, but other phenomenological parameters that describe PNJL model are set to those of SM QCD. We further considered several possible distinct phenomenological behaviors of the EoS in this model. We find that the resulting EoS behavior can be similar to that of the SM QCD transition, albeit centered around a higher critical temperature. 

We now discuss PBH production associated with a high temperature first order QCD transition, focusing on the regime above the electroweak scale.
Consider that inflation generically results in a broad, flat, primordial curvature (scalar) power spectrum that could arise in a broad class of models~\cite{Wands:1998yp,Leach:2000yw,Leach:2001zf,Biagetti:2018pjj}
\begin{equation}\label{eq:powerspec}
\mathcal{P}_{\zeta}(k)=A_s\Theta(k_{\rm max}-k)\Theta(k - k_{\rm min})~,
\end{equation}
where $k_{\rm max}$ is the cutoff scale, with $k_{\rm max} \gg k_{\rm min}$, $\Theta$ is the Heaviside step function and
$A_s$ is the amplitude. From $\mathcal{P}_{\zeta}(k)$ we can obtain the power spectrum of density perturbations $\mathcal{P}_{\delta}(k)$.

When a sufficiently large density fluctuation enters the Hubble horizon, PBHs can efficiently form from an overdense region. 
The total mass-energy in the causal horizon is~(e.g.~\cite{Sasaki:2018dmp, Carr:2020gox})
\begin{align} \label{eq:mhor}
    M_H \simeq&~ 12 M_\odot \left(\frac{k_{\ast}}{10^{6}\textrm{ Mpc}^{-1}}\right)^{-2}\left(\frac{g_\ast}{106.75}\right)^{-1/6} \\
    \simeq&~ 4.8 \times 10^{-10} M_{\odot} \Big(\dfrac{
    T}{10~\textrm{TeV}}\Big)^{-2} \left(\frac{g_\ast}{106.75}\right)^{-1/2}~, \notag
\end{align}
where 
$k_{\ast}$ is the comoving wavenumber corresponding to
the horizon length-scale at the epoch of black hole formation in the radiation-dominated era at temperature $T$, and $g_{\ast}$ accounts for the relativistic degrees
of freedom in the primordial plasma. In the second line of Eq.~\eqref{eq:mhor} we also have shown $M_H$ in terms of energy contained within the horizon
during the radiation dominated era. This indicates that typical masses of PBHs associated with high QCD transition temperature $T \gtrsim$~TeV could be well below the solar mass range associated with PBH formation at the SM QCD transition.

The fraction of energy density collapsing to PBHs at formation can be found from Press-Schechter formalism~\cite{Sasaki:2018dmp}
\begin{equation}
\label{eq:beta}
    \beta = 2 \int_{\delta_c}^{\infty} d\delta \frac{M_{\rm PBH}}{M_H} P(\delta, \sigma)~,
\end{equation}
where $\delta_c$ is the critical density contrast for collapse. The critical density $\delta_c$ depends on the EoS parameter $w$~\cite{Musco:2012au}. 
Here, $P(\delta,\sigma)$ is the probability distribution of density fluctuations entering the
horizon and is assumed here to be Gaussian,
$P(\delta,\sigma)=(1/\sigma)\sqrt{2/\pi}~\textrm{exp}[-\delta^2/2\sigma^2]$, with variance
\begin{equation}
    \sigma^2 = \int_{0}^{\infty} \mathcal{P}_\delta (k) W(kR)^2 {\cal{T}}^2(k,\eta) \frac{dk}{k}~,
\end{equation}
where $W(kR)= \textrm{exp}(-(kR)^2/4)$ is the Fourier transform of the window smoothing function over the horizon scale~\cite{Gow:2020bzo}, with $R \sim 1/k_{\ast}$ being the length scale of mode $k_{\ast}$ when it enters the horizon. The transfer function is a function of the conformal time $\eta$,
\begin{equation}
    {\cal{T}}(k,\eta) = 3\frac{\sin(k\eta/\sqrt{3})-(k\eta/\sqrt{3})\cos(k\eta/\sqrt{3})}{(k\eta/\sqrt{3})^3}~.
\end{equation}
 
We can employ Eq.~\ref{eq:beta} for the critical collapse method. First, we solve for density contrast $\delta^{\prime}$ at a given PBH mass and temperature 
\begin{equation}
\label{eq:deltaprime}
    \delta'(T) = \delta_c(T) + \left(\frac{M_{\rm PBH}}{K M_H(T)}\right)^{1/\gamma(T)}~,
\end{equation}
where $\gamma$ is the critical exponent that determines the scaling behavior of PBH mass and depends on $w(T)$, for which we follow the simulation results of Ref.~\cite{Musco:2012au}.
We can invert this to find $T(\delta')$ at given PBH mass. Since PBH with mass a $M_{\rm PBH}$ can be produced at multiple temperatures, we integrate over $T$ and impose a Dirac delta function $\delta(T-T(\delta'))$,
\begin{align}
\begin{split}
\label{eq:betamod}
    \beta(M)=&~2\int_{0}^{T_{\rm max}} dT \int_{\delta_c(T)}^{\infty} d\delta \frac{M}{M_H(T)}P(\delta)\delta(T-T(\delta')) \\
    =&~ 4\int_{0}^{T_{\rm max}} \frac{dT}{T} \frac{M}{M_H(T)}P(\delta'(T))(\delta'(T)-\delta_c(T))~.
\end{split}
\end{align}
Here $T_{\max}$ is the temperature at which $M_{\rm PBH} = M_H(T)$. In the second line we have swapped the order of integration and changed the Dirac delta variable from $T$ to $\delta$. From this modified $\beta$ function, the PBH fraction of DM is calculated as
\begin{equation}
\label{eq:fpbh}
    f_{\rm PBH} = \int \left(\frac{M}{M_{\rm eq}}\right)^{-1/2} \frac{\beta(M)}{\Omega_{\rm DM}} \frac{dM}{M}~,
\end{equation}
where $M_{\rm eq} = 3 \times 10^{17} M_{\odot}$ is the horizon mass at matter-radiation equality and $\Omega_{\rm DM}$ is the dark matter contribution to closure.
 
In Fig.~\ref{fig:PBHfull} we display results for our Model A, Model B and Model C descriptions of a high temperature QCD transition scenario (see Table I for details). The PBH mass spectra peak at the mass corresponding to the horizon mass $M_H(k)$ of Eq.~\eqref{eq:mhor} when the shortest wavelength $\sim 1/k_{\rm max}$ re-enters the horizon and has a tail of $\sim M^{-1/2}$, characteristic of broad power spectrum of Eq.~\eqref{eq:powerspec}~\cite{DeLuca:2020ioi}. On the other hand, a high temperature QCD transition results in a significantly pronounced peak at $M_{\rm PBH, peak}$ in the PBH mass distribution associated with horizon mass $M_H(T)$ of Eq.~\eqref{eq:mhor} around transition temperature $T_o$ (see Table I for model parameters). In Supplemental Material~\cite{supmat} we discuss how the PBH DM spectrum would look for a PNJL model as well as for other phenomenological approaches we consider.
Intriguingly, we find that a QCD transition at temperatures around $T \sim 200$~TeV could yield PBHs with masses that could account for all of the DM given current constraints. 
In contrast PBHs formed in an SM QCD scenario could constitute a sub-dominant DM component.~PBHs in this mass range are known to result in a  variety of intriguing observational signatures when they interact with neutron stars (e.g.~\cite{Fuller:2017uyd,Takhistov:2020vxs,Takhistov:2017nmt,Takhistov:2017bpt, Genolini:2020ejw, Dasgupta:2020mqg}).
We establish that optical telescopes conducting microlensing surveys, which have been shown to be excellent probes of PBHs~(e.g.~\cite{Niikura:2017zjd,Sugiyama:2019dgt,Kusenko:2020pcg,Sugiyama:2020roc}), could explore untested strong force regimes spanning decades in QCD transition temperatures above the electroweak scale. 

The Subaru Hyper-Suprime Cam (HSC) microlensing survey of the Andromeda galaxy (M31) reported a candidate event consistent with a PBH at $f_{\rm PBH} (M \sim 10^{-9} M_{\odot}) \sim 10^{-2}$~\cite{Niikura:2017zjd}.
We note that this detected HSC event is consistent with PBHs produced at a first order QCD transition around $T \sim 7$ TeV, as exemplified by Model B in Table I and Fig.~\ref{fig:PBHfull}. Intriguingly, the detected HSC event was obtained with only 7 hours of data. For reference, we also display Model C in Fig.~\ref{fig:PBHfull}, where the QCD transition occurs below the electroweak transition. While a detailed discussion of such a scenario is beyond the scope of this work, it exemplifies that a QCD transition around $T\sim 70$ GeV is consistent with the 6 Earth-mass candidate events detected in the 5-year survey of the Galactic bulge by the Optical Gravitational Lensing
Experiment (OGLE)~\cite{Niikura:2019kqi}. Future longer surveys could produce stronger constraints on PBH masses and contributions to closure and our work shows that these could have implications for beyond the SM extensions of QCD. In Supplemental Material~\cite{supmat} we demonstrate that the resulting abundance of PBHs $f_{\rm PBH}$ sensitively depends on the amplitude $A_s$ of the primordial power spectrum from Eq.~\eqref{eq:powerspec}. As induced GWs, as is clear in Eq.~\eqref{eq:inducedgw}, depend on $\sim A_s^2$, these signatures are also sensitive to variation in $A_s$.

Curvature perturbations resulting in PBHs could also lead to generation of induced gravitational waves at second order~\cite{Cai:2018dig,Ananda:2006af,Kohri:2018awv,Espinosa:2018eve,Domenech:2021ztg}. These could give a stochastic GW background (SGWB) at present with closure contribution
\begin{align} \label{eq:inducedgw}
    \Omega_{\rm GW} =&~ \frac{c_g \Omega_{r,0}}{972}\int_{0}^\infty dx \int_{|1-x|}^{1+x} dy \frac{x^2}{y^2} \left[1-\frac{(1+x^2-y^2)^2}{4x^2}\right]^2 \notag\\
    &~\times \mathcal{P}_{\zeta}(kx)\mathcal{P}_{\zeta}(ky)\mathcal{I}^2(x,y)~,
\end{align}
where $k = 2 \pi f$ and $f$ is the GW frequency, $c_g$ describes the change in the number of radiation degrees of freedom
over the evolution of the universe from the GW generation epoch to the present, $\Omega_{r,0}$ is the radiation contribution to closure today, and $\mathcal{I}(x,y)$ is the kernel function employed in the analytic solution obtained by Ref.~\cite{Espinosa:2018eve}.
In Fig.~\ref{fig:PBHfull} we display our resulting GW signatures and relevant observational limits. We note that GWs from Models A, B and C for PBHs produced from high temperature QCD phase transitions can account for the recently claimed signal from the 12.5 year analysis of the 
North American Nanohertz Observatory
for Gravitational Waves (NANOGrav) collaboration~\cite{NANOGrav:2020bcs}. These results will be tested with upcoming observations of LISA~\cite{LISA:2017} and the proposed $\mu$-Ares experiment~\cite{Sesana:2019vho}.
Ref.~\cite{DeLuca:2020agl,Sugiyama:2020roc} give a different 
context for the connection of the perturbation spectrum in Eq.~\eqref{eq:powerspec} with NANOGRav signatures.
Since PBH formation is exponentially sensitive to a small variation of the EoS, while the induced GWs are only linearly sensitive to it, we do not expect that effects from changes in the EoS parameter $w$ stemming from the QCD phase transition will significantly affect our results~(see also Ref.~\cite{Hajkarim:2019nbx}). 

In addition to induced GW signals, first order phase transitions are expected to also generate GW signatures (e.g.~\cite{Cutting:2018tjt}).
GWs of comoving frequency $f$ could result from a first order QCD transition at corresponding horizon mass 
$M_H \simeq 5.7 \times 10^{-10} M_{\odot} (10^{-4}~\textrm{Hz}/f)^2$. Hence, formation of asteroid-mass PBHs near the microlensing and open DM  window could be associated with observational signatures in upcoming GW observatories such as LISA~\cite{LISA:2017}, $\mu$-Ares~\cite{Sesana:2019vho} and DECIGO~\cite{Seto:2001qf}. This allows for an additional possibility for probing 
the considerations discussed here. Since production of such GWs strongly depends on complicated details of transition dynamics, we leave the analysis of such GW production coincident with PBH formation associated with high temperature QCD transition for future work.

In addition to the PBHs and GWs discussed above, further possible signatures could point to a high temperature QCD transition. These include potential high energy collider signals from a scalar $S$ coupled to gluons, and deviations from the standard Higgs couplings if that field mixes with the Higgs boson. Dynamics that restore the QCD transition to a conventional one occurring at low scales could  manifest in e.g. heavy ion collisions. Detailed discussion of these possibilities is beyond the scope of this work.

The novel connection between beyond SM QCD extensions, PBHs, and GWs is intriguing. It represents a promising connection between exciting particle physics possibilities and upcoming observations.
 
~\newline
\textit{Acknowledgements. - } We thank Seyda Ipek, Hitoshi Murayama, Ilia Musco, Misao Sasaki, Tim Tait, Graham White for helpful discussions. P.L. was supported by Grant Korea NRF2019R1C1C1010050.
V.T. acknowledges support by the World Premier International Research Center Initiative (WPI), MEXT, Japan and JSPS KAKENHI grant No. 23K13109. G.M.F.
acknowledges NSF Grant No. PHY-2209578 at UCSD
and the NSF N3AS Physics Frontier Center, NSF Grant
No. PHY-2020275, and the Heising-Simons Foundation
(2017-228).

 
\bibliography{bibliography}

 \newcommand{\noop}[1]{}
\begin{thebibliography}{99}%
\makeatletter
\providecommand \@ifxundefined [1]{%
 \@ifx{#1\undefined}
}%
\providecommand \@ifnum [1]{%
 \ifnum #1\expandafter \@firstoftwo
 \else \expandafter \@secondoftwo
 \fi
}%
\providecommand \@ifx [1]{%
 \ifx #1\expandafter \@firstoftwo
 \else \expandafter \@secondoftwo
 \fi
}%
\providecommand \natexlab [1]{#1}%
\providecommand \enquote  [1]{``#1''}%
\providecommand \bibnamefont  [1]{#1}%
\providecommand \bibfnamefont [1]{#1}%
\providecommand \citenamefont [1]{#1}%
\providecommand \href@noop [0]{\@secondoftwo}%
\providecommand \href [0]{\begingroup \@sanitize@url \@href}%
\providecommand \@href[1]{\@@startlink{#1}\@@href}%
\providecommand \@@href[1]{\endgroup#1\@@endlink}%
\providecommand \@sanitize@url [0]{\catcode `\\12\catcode `\$12\catcode
  `\&12\catcode `\#12\catcode `\^12\catcode `\_12\catcode `\%12\relax}%
\providecommand \@@startlink[1]{}%
\providecommand \@@endlink[0]{}%
\providecommand \url  [0]{\begingroup\@sanitize@url \@url }%
\providecommand \@url [1]{\endgroup\@href {#1}{\urlprefix }}%
\providecommand \urlprefix  [0]{URL }%
\providecommand \Eprint [0]{\href }%
\providecommand \doibase [0]{http://dx.doi.org/}%
\providecommand \selectlanguage [0]{\@gobble}%
\providecommand \bibinfo  [0]{\@secondoftwo}%
\providecommand \bibfield  [0]{\@secondoftwo}%
\providecommand \translation [1]{[#1]}%
\providecommand \BibitemOpen [0]{}%
\providecommand \bibitemStop [0]{}%
\providecommand \bibitemNoStop [0]{.\EOS\space}%
\providecommand \EOS [0]{\spacefactor3000\relax}%
\providecommand \BibitemShut  [1]{\csname bibitem#1\endcsname}%
\let\auto@bib@innerbib\@empty
\bibitem [{\citenamefont {Bhattacharya}\ \emph {et~al.}(2014)\citenamefont
  {Bhattacharya} \emph {et~al.}}]{Bhattacharya:2014ara}%
  \BibitemOpen
  \bibfield  {author} {\bibinfo {author} {\bibfnamefont {T.}~\bibnamefont
  {Bhattacharya}} \emph {et~al.},\ }\href {\doibase
  10.1103/PhysRevLett.113.082001} {\bibfield  {journal} {\bibinfo  {journal}
  {Phys. Rev. Lett.}\ }\textbf {\bibinfo {volume} {113}},\ \bibinfo {pages}
  {082001} (\bibinfo {year} {2014})},\ \Eprint {http://arxiv.org/abs/1402.5175}
  {arXiv:1402.5175 [hep-lat]} \BibitemShut {NoStop}%
\bibitem [{\citenamefont {Annala}\ \emph {et~al.}(2020)\citenamefont {Annala},
  \citenamefont {Gorda}, \citenamefont {Kurkela}, \citenamefont {N\"attil\"a},\
  and\ \citenamefont {Vuorinen}}]{Annala:2019puf}%
  \BibitemOpen
  \bibfield  {author} {\bibinfo {author} {\bibfnamefont {E.}~\bibnamefont
  {Annala}}, \bibinfo {author} {\bibfnamefont {T.}~\bibnamefont {Gorda}},
  \bibinfo {author} {\bibfnamefont {A.}~\bibnamefont {Kurkela}}, \bibinfo
  {author} {\bibfnamefont {J.}~\bibnamefont {N\"attil\"a}}, \ and\ \bibinfo
  {author} {\bibfnamefont {A.}~\bibnamefont {Vuorinen}},\ }\href {\doibase
  10.1038/s41567-020-0914-9} {\bibfield  {journal} {\bibinfo  {journal} {Nature
  Phys.}\ }\textbf {\bibinfo {volume} {16}},\ \bibinfo {pages} {907} (\bibinfo
  {year} {2020})},\ \Eprint {http://arxiv.org/abs/1903.09121} {arXiv:1903.09121
  [astro-ph.HE]} \BibitemShut {NoStop}%
\bibitem [{\citenamefont {Kawasaki}\ \emph {et~al.}(2000)\citenamefont
  {Kawasaki}, \citenamefont {Kohri},\ and\ \citenamefont
  {Sugiyama}}]{Kawasaki:2000en}%
  \BibitemOpen
  \bibfield  {author} {\bibinfo {author} {\bibfnamefont {M.}~\bibnamefont
  {Kawasaki}}, \bibinfo {author} {\bibfnamefont {K.}~\bibnamefont {Kohri}}, \
  and\ \bibinfo {author} {\bibfnamefont {N.}~\bibnamefont {Sugiyama}},\ }\href
  {\doibase 10.1103/PhysRevD.62.023506} {\bibfield  {journal} {\bibinfo
  {journal} {Phys. Rev. D}\ }\textbf {\bibinfo {volume} {62}},\ \bibinfo
  {pages} {023506} (\bibinfo {year} {2000})},\ \Eprint
  {http://arxiv.org/abs/astro-ph/0002127} {arXiv:astro-ph/0002127} \BibitemShut
  {NoStop}%
\bibitem [{\citenamefont {de~Salas}\ \emph {et~al.}(2015)\citenamefont
  {de~Salas}, \citenamefont {Lattanzi}, \citenamefont {Mangano}, \citenamefont
  {Miele}, \citenamefont {Pastor},\ and\ \citenamefont
  {Pisanti}}]{deSalas:2015glj}%
  \BibitemOpen
  \bibfield  {author} {\bibinfo {author} {\bibfnamefont {P.~F.}\ \bibnamefont
  {de~Salas}}, \bibinfo {author} {\bibfnamefont {M.}~\bibnamefont {Lattanzi}},
  \bibinfo {author} {\bibfnamefont {G.}~\bibnamefont {Mangano}}, \bibinfo
  {author} {\bibfnamefont {G.}~\bibnamefont {Miele}}, \bibinfo {author}
  {\bibfnamefont {S.}~\bibnamefont {Pastor}}, \ and\ \bibinfo {author}
  {\bibfnamefont {O.}~\bibnamefont {Pisanti}},\ }\href {\doibase
  10.1103/PhysRevD.92.123534} {\bibfield  {journal} {\bibinfo  {journal} {Phys.
  Rev.}\ }\textbf {\bibinfo {volume} {D92}},\ \bibinfo {pages} {123534}
  (\bibinfo {year} {2015})},\ \Eprint {http://arxiv.org/abs/1511.00672}
  {arXiv:1511.00672 [astro-ph.CO]} \BibitemShut {NoStop}%
\bibitem [{\citenamefont {Hasegawa}\ \emph {et~al.}(2019)\citenamefont
  {Hasegawa}, \citenamefont {Hiroshima}, \citenamefont {Kohri}, \citenamefont
  {Hansen}, \citenamefont {Tram},\ and\ \citenamefont
  {Hannestad}}]{Hasegawa:2019jsa}%
  \BibitemOpen
  \bibfield  {author} {\bibinfo {author} {\bibfnamefont {T.}~\bibnamefont
  {Hasegawa}}, \bibinfo {author} {\bibfnamefont {N.}~\bibnamefont {Hiroshima}},
  \bibinfo {author} {\bibfnamefont {K.}~\bibnamefont {Kohri}}, \bibinfo
  {author} {\bibfnamefont {R.~S.~L.}\ \bibnamefont {Hansen}}, \bibinfo {author}
  {\bibfnamefont {T.}~\bibnamefont {Tram}}, \ and\ \bibinfo {author}
  {\bibfnamefont {S.}~\bibnamefont {Hannestad}},\ }\href {\doibase
  10.1088/1475-7516/2019/12/012} {\bibfield  {journal} {\bibinfo  {journal}
  {JCAP}\ }\textbf {\bibinfo {volume} {12}},\ \bibinfo {pages} {012} (\bibinfo
  {year} {2019})},\ \Eprint {http://arxiv.org/abs/1908.10189} {arXiv:1908.10189
  [hep-ph]} \BibitemShut {NoStop}%
\bibitem [{\citenamefont {Fuller}\ \emph {et~al.}(1988)\citenamefont {Fuller},
  \citenamefont {Mathews},\ and\ \citenamefont {Alcock}}]{Fuller:1987ue}%
  \BibitemOpen
  \bibfield  {author} {\bibinfo {author} {\bibfnamefont {G.~M.}\ \bibnamefont
  {Fuller}}, \bibinfo {author} {\bibfnamefont {G.~J.}\ \bibnamefont {Mathews}},
  \ and\ \bibinfo {author} {\bibfnamefont {C.~R.}\ \bibnamefont {Alcock}},\
  }\href {\doibase 10.1103/PhysRevD.37.1380} {\bibfield  {journal} {\bibinfo
  {journal} {Phys. Rev. D}\ }\textbf {\bibinfo {volume} {37}},\ \bibinfo
  {pages} {1380} (\bibinfo {year} {1988})}\BibitemShut {NoStop}%
\bibitem [{\citenamefont {Byrnes}\ \emph {et~al.}(2018)\citenamefont {Byrnes},
  \citenamefont {Hindmarsh}, \citenamefont {Young},\ and\ \citenamefont
  {Hawkins}}]{Byrnes:2018clq}%
  \BibitemOpen
  \bibfield  {author} {\bibinfo {author} {\bibfnamefont {C.~T.}\ \bibnamefont
  {Byrnes}}, \bibinfo {author} {\bibfnamefont {M.}~\bibnamefont {Hindmarsh}},
  \bibinfo {author} {\bibfnamefont {S.}~\bibnamefont {Young}}, \ and\ \bibinfo
  {author} {\bibfnamefont {M.~R.~S.}\ \bibnamefont {Hawkins}},\ }\href
  {\doibase 10.1088/1475-7516/2018/08/041} {\bibfield  {journal} {\bibinfo
  {journal} {JCAP}\ }\textbf {\bibinfo {volume} {08}},\ \bibinfo {pages} {041}
  (\bibinfo {year} {2018})},\ \Eprint {http://arxiv.org/abs/1801.06138}
  {arXiv:1801.06138 [astro-ph.CO]} \BibitemShut {NoStop}%
\bibitem [{\citenamefont {Jedamzik}(1998)}]{Jedamzik:1998hc}%
  \BibitemOpen
  \bibfield  {author} {\bibinfo {author} {\bibfnamefont {K.}~\bibnamefont
  {Jedamzik}},\ }\href {\doibase 10.1016/S0370-1573(98)00067-2} {\bibfield
  {journal} {\bibinfo  {journal} {Phys. Rept.}\ }\textbf {\bibinfo {volume}
  {307}},\ \bibinfo {pages} {155} (\bibinfo {year} {1998})},\ \Eprint
  {http://arxiv.org/abs/astro-ph/9805147} {arXiv:astro-ph/9805147} \BibitemShut
  {NoStop}%
\bibitem [{\citenamefont {Jedamzik}(2021)}]{Jedamzik:2020omx}%
  \BibitemOpen
  \bibfield  {author} {\bibinfo {author} {\bibfnamefont {K.}~\bibnamefont
  {Jedamzik}},\ }\href {\doibase 10.1103/PhysRevLett.126.051302} {\bibfield
  {journal} {\bibinfo  {journal} {Phys. Rev. Lett.}\ }\textbf {\bibinfo
  {volume} {126}},\ \bibinfo {pages} {051302} (\bibinfo {year} {2021})},\
  \Eprint {http://arxiv.org/abs/2007.03565} {arXiv:2007.03565 [astro-ph.CO]}
  \BibitemShut {NoStop}%
\bibitem [{\citenamefont {Carr}\ \emph
  {et~al.}(2021{\natexlab{a}})\citenamefont {Carr}, \citenamefont {Clesse},
  \citenamefont {Garc\'\i{}a-Bellido},\ and\ \citenamefont
  {K\"uhnel}}]{Carr:2019kxo}%
  \BibitemOpen
  \bibfield  {author} {\bibinfo {author} {\bibfnamefont {B.}~\bibnamefont
  {Carr}}, \bibinfo {author} {\bibfnamefont {S.}~\bibnamefont {Clesse}},
  \bibinfo {author} {\bibfnamefont {J.}~\bibnamefont {Garc\'\i{}a-Bellido}}, \
  and\ \bibinfo {author} {\bibfnamefont {F.}~\bibnamefont {K\"uhnel}},\ }\href
  {\doibase 10.1016/j.dark.2020.100755} {\bibfield  {journal} {\bibinfo
  {journal} {Phys. Dark Univ.}\ }\textbf {\bibinfo {volume} {31}},\ \bibinfo
  {pages} {100755} (\bibinfo {year} {2021}{\natexlab{a}})},\ \Eprint
  {http://arxiv.org/abs/1906.08217} {arXiv:1906.08217 [astro-ph.CO]}
  \BibitemShut {NoStop}%
\bibitem [{\citenamefont {Carr}\ \emph
  {et~al.}(2021{\natexlab{b}})\citenamefont {Carr}, \citenamefont {Clesse},\
  and\ \citenamefont {Garc\'\i{}a-Bellido}}]{Carr:2019hud}%
  \BibitemOpen
  \bibfield  {author} {\bibinfo {author} {\bibfnamefont {B.}~\bibnamefont
  {Carr}}, \bibinfo {author} {\bibfnamefont {S.}~\bibnamefont {Clesse}}, \ and\
  \bibinfo {author} {\bibfnamefont {J.}~\bibnamefont {Garc\'\i{}a-Bellido}},\
  }\href {\doibase 10.1093/mnras/staa3726} {\bibfield  {journal} {\bibinfo
  {journal} {Mon. Not. Roy. Astron. Soc.}\ }\textbf {\bibinfo {volume} {501}},\
  \bibinfo {pages} {1426} (\bibinfo {year} {2021}{\natexlab{b}})},\ \Eprint
  {http://arxiv.org/abs/1904.02129} {arXiv:1904.02129 [astro-ph.CO]}
  \BibitemShut {NoStop}%
\bibitem [{\citenamefont {Dolgov}\ and\ \citenamefont
  {Postnov}(2020)}]{Dolgov:2020sov}%
  \BibitemOpen
  \bibfield  {author} {\bibinfo {author} {\bibfnamefont {A.}~\bibnamefont
  {Dolgov}}\ and\ \bibinfo {author} {\bibfnamefont {K.}~\bibnamefont
  {Postnov}},\ }\href {\doibase 10.1088/1475-7516/2020/07/063} {\bibfield
  {journal} {\bibinfo  {journal} {JCAP}\ }\textbf {\bibinfo {volume} {07}},\
  \bibinfo {pages} {063} (\bibinfo {year} {2020})},\ \Eprint
  {http://arxiv.org/abs/2004.11669} {arXiv:2004.11669 [astro-ph.CO]}
  \BibitemShut {NoStop}%
\bibitem [{\citenamefont {Franciolini}\ \emph
  {et~al.}(2022{\natexlab{a}})\citenamefont {Franciolini}, \citenamefont
  {Musco}, \citenamefont {Pani},\ and\ \citenamefont
  {Urbano}}]{Franciolini:2022tfm}%
  \BibitemOpen
  \bibfield  {author} {\bibinfo {author} {\bibfnamefont {G.}~\bibnamefont
  {Franciolini}}, \bibinfo {author} {\bibfnamefont {I.}~\bibnamefont {Musco}},
  \bibinfo {author} {\bibfnamefont {P.}~\bibnamefont {Pani}}, \ and\ \bibinfo
  {author} {\bibfnamefont {A.}~\bibnamefont {Urbano}},\ }\href@noop {} {\
  (\bibinfo {year} {2022}{\natexlab{a}})},\ \Eprint
  {http://arxiv.org/abs/2209.05959} {arXiv:2209.05959 [astro-ph.CO]}
  \BibitemShut {NoStop}%
\bibitem [{\citenamefont {Juan}\ \emph {et~al.}(2022)\citenamefont {Juan},
  \citenamefont {Serpico},\ and\ \citenamefont
  {Franco~Abell\'an}}]{Juan:2022mir}%
  \BibitemOpen
  \bibfield  {author} {\bibinfo {author} {\bibfnamefont {J.~I.}\ \bibnamefont
  {Juan}}, \bibinfo {author} {\bibfnamefont {P.~D.}\ \bibnamefont {Serpico}}, \
  and\ \bibinfo {author} {\bibfnamefont {G.}~\bibnamefont {Franco~Abell\'an}},\
  }\href {\doibase 10.1088/1475-7516/2022/07/009} {\bibfield  {journal}
  {\bibinfo  {journal} {JCAP}\ }\textbf {\bibinfo {volume} {07}},\ \bibinfo
  {pages} {009} (\bibinfo {year} {2022})},\ \Eprint
  {http://arxiv.org/abs/2204.07027} {arXiv:2204.07027 [astro-ph.CO]}
  \BibitemShut {NoStop}%
\bibitem [{\citenamefont {Clesse}\ and\ \citenamefont
  {Garcia-Bellido}(2020)}]{Clesse:2020ghq}%
  \BibitemOpen
  \bibfield  {author} {\bibinfo {author} {\bibfnamefont {S.}~\bibnamefont
  {Clesse}}\ and\ \bibinfo {author} {\bibfnamefont {J.}~\bibnamefont
  {Garcia-Bellido}},\ }\href@noop {} {\  (\bibinfo {year} {2020})},\ \Eprint
  {http://arxiv.org/abs/2007.06481} {arXiv:2007.06481 [astro-ph.CO]}
  \BibitemShut {NoStop}%
\bibitem [{\citenamefont {{Zel'dovich}}\ and\ \citenamefont
  {{Novikov}}(1967)}]{Zeldovich:1967}%
  \BibitemOpen
  \bibfield  {author} {\bibinfo {author} {\bibfnamefont {Y.~B.}\ \bibnamefont
  {{Zel'dovich}}}\ and\ \bibinfo {author} {\bibfnamefont {I.~D.}\ \bibnamefont
  {{Novikov}}},\ }\href@noop {} {\bibfield  {journal} {\bibinfo  {journal}
  {Sov. Astron.}\ }\textbf {\bibinfo {volume} {10}},\ \bibinfo {pages} {602}
  (\bibinfo {year} {1967})}\BibitemShut {NoStop}%
\bibitem [{\citenamefont {Hawking}(1971)}]{Hawking:1971ei}%
  \BibitemOpen
  \bibfield  {author} {\bibinfo {author} {\bibfnamefont {S.}~\bibnamefont
  {Hawking}},\ }\href@noop {} {\bibfield  {journal} {\bibinfo  {journal} {Mon.
  Not. Roy. Astron. Soc.}\ }\textbf {\bibinfo {volume} {152}},\ \bibinfo
  {pages} {75} (\bibinfo {year} {1971})}\BibitemShut {NoStop}%
\bibitem [{\citenamefont {Carr}\ and\ \citenamefont
  {Hawking}(1974)}]{Carr:1974nx}%
  \BibitemOpen
  \bibfield  {author} {\bibinfo {author} {\bibfnamefont {B.~J.}\ \bibnamefont
  {Carr}}\ and\ \bibinfo {author} {\bibfnamefont {S.~W.}\ \bibnamefont
  {Hawking}},\ }\href@noop {} {\bibfield  {journal} {\bibinfo  {journal} {Mon.
  Not. Roy. Astron. Soc.}\ }\textbf {\bibinfo {volume} {168}},\ \bibinfo
  {pages} {399} (\bibinfo {year} {1974})}\BibitemShut {NoStop}%
\bibitem [{\citenamefont {Garcia-Bellido}\ \emph {et~al.}(1996)\citenamefont
  {Garcia-Bellido}, \citenamefont {Linde},\ and\ \citenamefont
  {Wands}}]{GarciaBellido:1996qt}%
  \BibitemOpen
  \bibfield  {author} {\bibinfo {author} {\bibfnamefont {J.}~\bibnamefont
  {Garcia-Bellido}}, \bibinfo {author} {\bibfnamefont {A.~D.}\ \bibnamefont
  {Linde}}, \ and\ \bibinfo {author} {\bibfnamefont {D.}~\bibnamefont
  {Wands}},\ }\href {\doibase 10.1103/PhysRevD.54.6040} {\bibfield  {journal}
  {\bibinfo  {journal} {Phys. Rev.}\ }\textbf {\bibinfo {volume} {D54}},\
  \bibinfo {pages} {6040} (\bibinfo {year} {1996})},\ \Eprint
  {http://arxiv.org/abs/astro-ph/9605094} {arXiv:astro-ph/9605094 [astro-ph]}
  \BibitemShut {NoStop}%
\bibitem [{\citenamefont {Green}\ and\ \citenamefont
  {Malik}(2001)}]{Green:2000he}%
  \BibitemOpen
  \bibfield  {author} {\bibinfo {author} {\bibfnamefont {A.~M.}\ \bibnamefont
  {Green}}\ and\ \bibinfo {author} {\bibfnamefont {K.~A.}\ \bibnamefont
  {Malik}},\ }\href {\doibase 10.1103/PhysRevD.64.021301} {\bibfield  {journal}
  {\bibinfo  {journal} {Phys. Rev. D}\ }\textbf {\bibinfo {volume} {64}},\
  \bibinfo {pages} {021301} (\bibinfo {year} {2001})},\ \Eprint
  {http://arxiv.org/abs/hep-ph/0008113} {arXiv:hep-ph/0008113} \BibitemShut
  {NoStop}%
\bibitem [{\citenamefont {Khlopov}(2010)}]{Khlopov:2008qy}%
  \BibitemOpen
  \bibfield  {author} {\bibinfo {author} {\bibfnamefont {M.~{\relax Yu}.}\
  \bibnamefont {Khlopov}},\ }\href {\doibase 10.1088/1674-4527/10/6/001}
  {\bibfield  {journal} {\bibinfo  {journal} {Res. Astron. Astrophys.}\
  }\textbf {\bibinfo {volume} {10}},\ \bibinfo {pages} {495} (\bibinfo {year}
  {2010})},\ \Eprint {http://arxiv.org/abs/0801.0116} {arXiv:0801.0116
  [astro-ph]} \BibitemShut {NoStop}%
\bibitem [{\citenamefont {Frampton}\ \emph {et~al.}(2010)\citenamefont
  {Frampton}, \citenamefont {Kawasaki}, \citenamefont {Takahashi},\ and\
  \citenamefont {Yanagida}}]{Frampton:2010sw}%
  \BibitemOpen
  \bibfield  {author} {\bibinfo {author} {\bibfnamefont {P.~H.}\ \bibnamefont
  {Frampton}}, \bibinfo {author} {\bibfnamefont {M.}~\bibnamefont {Kawasaki}},
  \bibinfo {author} {\bibfnamefont {F.}~\bibnamefont {Takahashi}}, \ and\
  \bibinfo {author} {\bibfnamefont {T.~T.}\ \bibnamefont {Yanagida}},\ }\href
  {\doibase 10.1088/1475-7516/2010/04/023} {\bibfield  {journal} {\bibinfo
  {journal} {JCAP}\ }\textbf {\bibinfo {volume} {1004}},\ \bibinfo {pages}
  {023} (\bibinfo {year} {2010})},\ \Eprint {http://arxiv.org/abs/1001.2308}
  {arXiv:1001.2308 [hep-ph]} \BibitemShut {NoStop}%
\bibitem [{\citenamefont {Cotner}\ \emph {et~al.}(2019)\citenamefont {Cotner},
  \citenamefont {Kusenko}, \citenamefont {Sasaki},\ and\ \citenamefont
  {Takhistov}}]{Cotner:2019ykd}%
  \BibitemOpen
  \bibfield  {author} {\bibinfo {author} {\bibfnamefont {E.}~\bibnamefont
  {Cotner}}, \bibinfo {author} {\bibfnamefont {A.}~\bibnamefont {Kusenko}},
  \bibinfo {author} {\bibfnamefont {M.}~\bibnamefont {Sasaki}}, \ and\ \bibinfo
  {author} {\bibfnamefont {V.}~\bibnamefont {Takhistov}},\ }\href {\doibase
  10.1088/1475-7516/2019/10/077} {\bibfield  {journal} {\bibinfo  {journal}
  {JCAP}\ }\textbf {\bibinfo {volume} {1910}},\ \bibinfo {pages} {077}
  (\bibinfo {year} {2019})},\ \Eprint {http://arxiv.org/abs/1907.10613}
  {arXiv:1907.10613 [astro-ph.CO]} \BibitemShut {NoStop}%
\bibitem [{\citenamefont {Cotner}\ \emph {et~al.}(2018)\citenamefont {Cotner},
  \citenamefont {Kusenko},\ and\ \citenamefont {Takhistov}}]{Cotner:2018vug}%
  \BibitemOpen
  \bibfield  {author} {\bibinfo {author} {\bibfnamefont {E.}~\bibnamefont
  {Cotner}}, \bibinfo {author} {\bibfnamefont {A.}~\bibnamefont {Kusenko}}, \
  and\ \bibinfo {author} {\bibfnamefont {V.}~\bibnamefont {Takhistov}},\ }\href
  {\doibase 10.1103/PhysRevD.98.083513} {\bibfield  {journal} {\bibinfo
  {journal} {Phys. Rev.}\ }\textbf {\bibinfo {volume} {D98}},\ \bibinfo {pages}
  {083513} (\bibinfo {year} {2018})},\ \Eprint
  {http://arxiv.org/abs/1801.03321} {arXiv:1801.03321 [astro-ph.CO]}
  \BibitemShut {NoStop}%
\bibitem [{\citenamefont {Green}(2016)}]{Green:2016xgy}%
  \BibitemOpen
  \bibfield  {author} {\bibinfo {author} {\bibfnamefont {A.~M.}\ \bibnamefont
  {Green}},\ }\href {\doibase 10.1103/PhysRevD.94.063530} {\bibfield  {journal}
  {\bibinfo  {journal} {Phys. Rev. D}\ }\textbf {\bibinfo {volume} {94}},\
  \bibinfo {pages} {063530} (\bibinfo {year} {2016})},\ \Eprint
  {http://arxiv.org/abs/1609.01143} {arXiv:1609.01143 [astro-ph.CO]}
  \BibitemShut {NoStop}%
\bibitem [{\citenamefont {Kusenko}\ \emph {et~al.}(2020)\citenamefont
  {Kusenko}, \citenamefont {Sasaki}, \citenamefont {Sugiyama}, \citenamefont
  {Takada}, \citenamefont {Takhistov},\ and\ \citenamefont
  {Vitagliano}}]{Kusenko:2020pcg}%
  \BibitemOpen
  \bibfield  {author} {\bibinfo {author} {\bibfnamefont {A.}~\bibnamefont
  {Kusenko}}, \bibinfo {author} {\bibfnamefont {M.}~\bibnamefont {Sasaki}},
  \bibinfo {author} {\bibfnamefont {S.}~\bibnamefont {Sugiyama}}, \bibinfo
  {author} {\bibfnamefont {M.}~\bibnamefont {Takada}}, \bibinfo {author}
  {\bibfnamefont {V.}~\bibnamefont {Takhistov}}, \ and\ \bibinfo {author}
  {\bibfnamefont {E.}~\bibnamefont {Vitagliano}},\ }\href@noop {} {\  (\bibinfo
  {year} {2020})},\ \Eprint {http://arxiv.org/abs/2001.09160} {arXiv:2001.09160
  [astro-ph.CO]} \BibitemShut {NoStop}%
\bibitem [{\citenamefont {Sasaki}\ \emph {et~al.}(2018)\citenamefont {Sasaki},
  \citenamefont {Suyama}, \citenamefont {Tanaka},\ and\ \citenamefont
  {Yokoyama}}]{Sasaki:2018dmp}%
  \BibitemOpen
  \bibfield  {author} {\bibinfo {author} {\bibfnamefont {M.}~\bibnamefont
  {Sasaki}}, \bibinfo {author} {\bibfnamefont {T.}~\bibnamefont {Suyama}},
  \bibinfo {author} {\bibfnamefont {T.}~\bibnamefont {Tanaka}}, \ and\ \bibinfo
  {author} {\bibfnamefont {S.}~\bibnamefont {Yokoyama}},\ }\href {\doibase
  10.1088/1361-6382/aaa7b4} {\bibfield  {journal} {\bibinfo  {journal} {Class.
  Quant. Grav.}\ }\textbf {\bibinfo {volume} {35}},\ \bibinfo {pages} {063001}
  (\bibinfo {year} {2018})},\ \Eprint {http://arxiv.org/abs/1801.05235}
  {arXiv:1801.05235 [astro-ph.CO]} \BibitemShut {NoStop}%
\bibitem [{\citenamefont {Carr}\ \emph {et~al.}(2020)\citenamefont {Carr},
  \citenamefont {Kohri}, \citenamefont {Sendouda},\ and\ \citenamefont
  {Yokoyama}}]{Carr:2020gox}%
  \BibitemOpen
  \bibfield  {author} {\bibinfo {author} {\bibfnamefont {B.}~\bibnamefont
  {Carr}}, \bibinfo {author} {\bibfnamefont {K.}~\bibnamefont {Kohri}},
  \bibinfo {author} {\bibfnamefont {Y.}~\bibnamefont {Sendouda}}, \ and\
  \bibinfo {author} {\bibfnamefont {J.}~\bibnamefont {Yokoyama}},\ }\href@noop
  {} {\  (\bibinfo {year} {2020})},\ \Eprint {http://arxiv.org/abs/2002.12778}
  {arXiv:2002.12778 [astro-ph.CO]} \BibitemShut {NoStop}%
\bibitem [{\citenamefont {Green}\ and\ \citenamefont
  {Kavanagh}(2020)}]{Green:2020jor}%
  \BibitemOpen
  \bibfield  {author} {\bibinfo {author} {\bibfnamefont {A.~M.}\ \bibnamefont
  {Green}}\ and\ \bibinfo {author} {\bibfnamefont {B.~J.}\ \bibnamefont
  {Kavanagh}},\ }\href@noop {} {\  (\bibinfo {year} {2020})},\ \Eprint
  {http://arxiv.org/abs/2007.10722} {arXiv:2007.10722 [astro-ph.CO]}
  \BibitemShut {NoStop}%
\bibitem [{\citenamefont {Escriva}\ \emph {et~al.}(2022)\citenamefont
  {Escriva}, \citenamefont {Kuhnel},\ and\ \citenamefont
  {Tada}}]{Escriva:2022duf}%
  \BibitemOpen
  \bibfield  {author} {\bibinfo {author} {\bibfnamefont {A.}~\bibnamefont
  {Escriva}}, \bibinfo {author} {\bibfnamefont {F.}~\bibnamefont {Kuhnel}}, \
  and\ \bibinfo {author} {\bibfnamefont {Y.}~\bibnamefont {Tada}},\ }\href@noop
  {} {\  (\bibinfo {year} {2022})},\ \Eprint {http://arxiv.org/abs/2211.05767}
  {arXiv:2211.05767 [astro-ph.CO]} \BibitemShut {NoStop}%
\bibitem [{\citenamefont {Nakamura}\ \emph {et~al.}(1997)\citenamefont
  {Nakamura}, \citenamefont {Sasaki}, \citenamefont {Tanaka},\ and\
  \citenamefont {Thorne}}]{Nakamura:1997sm}%
  \BibitemOpen
  \bibfield  {author} {\bibinfo {author} {\bibfnamefont {T.}~\bibnamefont
  {Nakamura}}, \bibinfo {author} {\bibfnamefont {M.}~\bibnamefont {Sasaki}},
  \bibinfo {author} {\bibfnamefont {T.}~\bibnamefont {Tanaka}}, \ and\ \bibinfo
  {author} {\bibfnamefont {K.~S.}\ \bibnamefont {Thorne}},\ }\href {\doibase
  10.1086/310886} {\bibfield  {journal} {\bibinfo  {journal} {Astrophys. J.}\
  }\textbf {\bibinfo {volume} {487}},\ \bibinfo {pages} {L139} (\bibinfo {year}
  {1997})},\ \Eprint {http://arxiv.org/abs/astro-ph/9708060}
  {arXiv:astro-ph/9708060 [astro-ph]} \BibitemShut {NoStop}%
\bibitem [{\citenamefont {Clesse}\ and\ \citenamefont
  {Garcia-Bellido}(2015)}]{Clesse:2015wea}%
  \BibitemOpen
  \bibfield  {author} {\bibinfo {author} {\bibfnamefont {S.}~\bibnamefont
  {Clesse}}\ and\ \bibinfo {author} {\bibfnamefont {J.}~\bibnamefont
  {Garcia-Bellido}},\ }\href {\doibase 10.1103/PhysRevD.92.023524} {\bibfield
  {journal} {\bibinfo  {journal} {Phys. Rev.}\ }\textbf {\bibinfo {volume}
  {D92}},\ \bibinfo {pages} {023524} (\bibinfo {year} {2015})},\ \Eprint
  {http://arxiv.org/abs/1501.07565} {arXiv:1501.07565 [astro-ph.CO]}
  \BibitemShut {NoStop}%
\bibitem [{\citenamefont {Bird}\ \emph {et~al.}(2016)\citenamefont {Bird} \emph
  {et~al.}}]{Bird:2016dcv}%
  \BibitemOpen
  \bibfield  {author} {\bibinfo {author} {\bibfnamefont {S.}~\bibnamefont
  {Bird}} \emph {et~al.},\ }\href {\doibase 10.1103/PhysRevLett.116.201301}
  {\bibfield  {journal} {\bibinfo  {journal} {Phys. Rev. Lett.}\ }\textbf
  {\bibinfo {volume} {116}},\ \bibinfo {pages} {201301} (\bibinfo {year}
  {2016})},\ \Eprint {http://arxiv.org/abs/1603.00464} {arXiv:1603.00464
  [astro-ph.CO]} \BibitemShut {NoStop}%
\bibitem [{\citenamefont {Raidal}\ \emph {et~al.}(2017)\citenamefont {Raidal},
  \citenamefont {Vaskonen},\ and\ \citenamefont {Veermae}}]{Raidal:2017mfl}%
  \BibitemOpen
  \bibfield  {author} {\bibinfo {author} {\bibfnamefont {M.}~\bibnamefont
  {Raidal}}, \bibinfo {author} {\bibfnamefont {V.}~\bibnamefont {Vaskonen}}, \
  and\ \bibinfo {author} {\bibfnamefont {H.}~\bibnamefont {Veermae}},\
  }\href@noop {} {\  (\bibinfo {year} {2017})},\ \Eprint
  {http://arxiv.org/abs/1707.01480} {arXiv:1707.01480 [astro-ph.CO]}
  \BibitemShut {NoStop}%
\bibitem [{\citenamefont {Eroshenko}(2016)}]{Eroshenko:2016hmn}%
  \BibitemOpen
  \bibfield  {author} {\bibinfo {author} {\bibfnamefont {{\relax Yu}.~N.}\
  \bibnamefont {Eroshenko}},\ }\href@noop {} {\  (\bibinfo {year} {2016})},\
  \Eprint {http://arxiv.org/abs/1604.04932} {arXiv:1604.04932 [astro-ph.CO]}
  \BibitemShut {NoStop}%
\bibitem [{\citenamefont {Sasaki}\ \emph {et~al.}(2016)\citenamefont {Sasaki},
  \citenamefont {Suyama}, \citenamefont {Tanaka},\ and\ \citenamefont
  {Yokoyama}}]{Sasaki:2016jop}%
  \BibitemOpen
  \bibfield  {author} {\bibinfo {author} {\bibfnamefont {M.}~\bibnamefont
  {Sasaki}}, \bibinfo {author} {\bibfnamefont {T.}~\bibnamefont {Suyama}},
  \bibinfo {author} {\bibfnamefont {T.}~\bibnamefont {Tanaka}}, \ and\ \bibinfo
  {author} {\bibfnamefont {S.}~\bibnamefont {Yokoyama}},\ }\href {\doibase
  10.1103/PhysRevLett.117.061101} {\bibfield  {journal} {\bibinfo  {journal}
  {Phys. Rev. Lett.}\ }\textbf {\bibinfo {volume} {117}},\ \bibinfo {pages}
  {061101} (\bibinfo {year} {2016})},\ \Eprint
  {http://arxiv.org/abs/1603.08338} {arXiv:1603.08338 [astro-ph.CO]}
  \BibitemShut {NoStop}%
\bibitem [{\citenamefont {Clesse}\ and\ \citenamefont
  {Garcia-Bellido}(2016)}]{Clesse:2016ajp}%
  \BibitemOpen
  \bibfield  {author} {\bibinfo {author} {\bibfnamefont {S.}~\bibnamefont
  {Clesse}}\ and\ \bibinfo {author} {\bibfnamefont {J.}~\bibnamefont
  {Garcia-Bellido}},\ }\href@noop {} {\  (\bibinfo {year} {2016})},\ \Eprint
  {http://arxiv.org/abs/1610.08479} {arXiv:1610.08479 [astro-ph.CO]}
  \BibitemShut {NoStop}%
\bibitem [{\citenamefont {Fuller}\ \emph {et~al.}(2017)\citenamefont {Fuller},
  \citenamefont {Kusenko},\ and\ \citenamefont {Takhistov}}]{Fuller:2017uyd}%
  \BibitemOpen
  \bibfield  {author} {\bibinfo {author} {\bibfnamefont {G.~M.}\ \bibnamefont
  {Fuller}}, \bibinfo {author} {\bibfnamefont {A.}~\bibnamefont {Kusenko}}, \
  and\ \bibinfo {author} {\bibfnamefont {V.}~\bibnamefont {Takhistov}},\ }\href
  {\doibase 10.1103/PhysRevLett.119.061101} {\bibfield  {journal} {\bibinfo
  {journal} {Phys. Rev. Lett.}\ }\textbf {\bibinfo {volume} {119}},\ \bibinfo
  {pages} {061101} (\bibinfo {year} {2017})},\ \Eprint
  {http://arxiv.org/abs/1704.01129} {arXiv:1704.01129 [astro-ph.HE]}
  \BibitemShut {NoStop}%
\bibitem [{\citenamefont {Takhistov}(2019)}]{Takhistov:2017nmt}%
  \BibitemOpen
  \bibfield  {author} {\bibinfo {author} {\bibfnamefont {V.}~\bibnamefont
  {Takhistov}},\ }\href {\doibase 10.1016/j.physletb.2018.12.043} {\bibfield
  {journal} {\bibinfo  {journal} {Phys. Lett.}\ }\textbf {\bibinfo {volume}
  {B789}},\ \bibinfo {pages} {538} (\bibinfo {year} {2019})},\ \Eprint
  {http://arxiv.org/abs/1710.09458} {arXiv:1710.09458 [astro-ph.HE]}
  \BibitemShut {NoStop}%
\bibitem [{\citenamefont {Takhistov}(2018)}]{Takhistov:2017bpt}%
  \BibitemOpen
  \bibfield  {author} {\bibinfo {author} {\bibfnamefont {V.}~\bibnamefont
  {Takhistov}},\ }\href {\doibase 10.1016/j.physletb.2018.05.026} {\bibfield
  {journal} {\bibinfo  {journal} {Phys. Lett.}\ }\textbf {\bibinfo {volume}
  {B782}},\ \bibinfo {pages} {77} (\bibinfo {year} {2018})},\ \Eprint
  {http://arxiv.org/abs/1707.05849} {arXiv:1707.05849 [astro-ph.CO]}
  \BibitemShut {NoStop}%
\bibitem [{\citenamefont {Franciolini}\ \emph
  {et~al.}(2022{\natexlab{b}})\citenamefont {Franciolini}, \citenamefont
  {Baibhav}, \citenamefont {De~Luca}, \citenamefont {Ng}, \citenamefont {Wong},
  \citenamefont {Berti}, \citenamefont {Pani}, \citenamefont {Riotto},\ and\
  \citenamefont {Vitale}}]{Franciolini:2021tla}%
  \BibitemOpen
  \bibfield  {author} {\bibinfo {author} {\bibfnamefont {G.}~\bibnamefont
  {Franciolini}}, \bibinfo {author} {\bibfnamefont {V.}~\bibnamefont
  {Baibhav}}, \bibinfo {author} {\bibfnamefont {V.}~\bibnamefont {De~Luca}},
  \bibinfo {author} {\bibfnamefont {K.~K.~Y.}\ \bibnamefont {Ng}}, \bibinfo
  {author} {\bibfnamefont {K.~W.~K.}\ \bibnamefont {Wong}}, \bibinfo {author}
  {\bibfnamefont {E.}~\bibnamefont {Berti}}, \bibinfo {author} {\bibfnamefont
  {P.}~\bibnamefont {Pani}}, \bibinfo {author} {\bibfnamefont {A.}~\bibnamefont
  {Riotto}}, \ and\ \bibinfo {author} {\bibfnamefont {S.}~\bibnamefont
  {Vitale}},\ }\href {\doibase 10.1103/PhysRevD.105.083526} {\bibfield
  {journal} {\bibinfo  {journal} {Phys. Rev. D}\ }\textbf {\bibinfo {volume}
  {105}},\ \bibinfo {pages} {083526} (\bibinfo {year} {2022}{\natexlab{b}})},\
  \Eprint {http://arxiv.org/abs/2105.03349} {arXiv:2105.03349 [gr-qc]}
  \BibitemShut {NoStop}%
\bibitem [{\citenamefont {Pisarski}\ and\ \citenamefont
  {Wilczek}(1984)}]{Pisarski:1983ms}%
  \BibitemOpen
  \bibfield  {author} {\bibinfo {author} {\bibfnamefont {R.~D.}\ \bibnamefont
  {Pisarski}}\ and\ \bibinfo {author} {\bibfnamefont {F.}~\bibnamefont
  {Wilczek}},\ }\href {\doibase 10.1103/PhysRevD.29.338} {\bibfield  {journal}
  {\bibinfo  {journal} {Phys. Rev. D}\ }\textbf {\bibinfo {volume} {29}},\
  \bibinfo {pages} {338} (\bibinfo {year} {1984})}\BibitemShut {NoStop}%
\bibitem [{\citenamefont {Davoudiasl}(2019)}]{Davoudiasl:2019ugw}%
  \BibitemOpen
  \bibfield  {author} {\bibinfo {author} {\bibfnamefont {H.}~\bibnamefont
  {Davoudiasl}},\ }\href {\doibase 10.1103/PhysRevLett.123.101102} {\bibfield
  {journal} {\bibinfo  {journal} {Phys. Rev. Lett.}\ }\textbf {\bibinfo
  {volume} {123}},\ \bibinfo {pages} {101102} (\bibinfo {year} {2019})},\
  \Eprint {http://arxiv.org/abs/1902.07805} {arXiv:1902.07805 [hep-ph]}
  \BibitemShut {NoStop}%
\bibitem [{\citenamefont {Gross}\ \emph {et~al.}(2021)\citenamefont {Gross},
  \citenamefont {Landini}, \citenamefont {Strumia},\ and\ \citenamefont
  {Teresi}}]{Gross:2021qgx}%
  \BibitemOpen
  \bibfield  {author} {\bibinfo {author} {\bibfnamefont {C.}~\bibnamefont
  {Gross}}, \bibinfo {author} {\bibfnamefont {G.}~\bibnamefont {Landini}},
  \bibinfo {author} {\bibfnamefont {A.}~\bibnamefont {Strumia}}, \ and\
  \bibinfo {author} {\bibfnamefont {D.}~\bibnamefont {Teresi}},\ }\href
  {\doibase 10.1007/JHEP09(2021)033} {\bibfield  {journal} {\bibinfo  {journal}
  {JHEP}\ }\textbf {\bibinfo {volume} {09}},\ \bibinfo {pages} {033} (\bibinfo
  {year} {2021})},\ \Eprint {http://arxiv.org/abs/2105.02840} {arXiv:2105.02840
  [hep-ph]} \BibitemShut {NoStop}%
\bibitem [{\citenamefont {Ipek}\ and\ \citenamefont
  {Tait}(2019)}]{Ipek:2018lhm}%
  \BibitemOpen
  \bibfield  {author} {\bibinfo {author} {\bibfnamefont {S.}~\bibnamefont
  {Ipek}}\ and\ \bibinfo {author} {\bibfnamefont {T.~M.~P.}\ \bibnamefont
  {Tait}},\ }\href {\doibase 10.1103/PhysRevLett.122.112001} {\bibfield
  {journal} {\bibinfo  {journal} {Phys. Rev. Lett.}\ }\textbf {\bibinfo
  {volume} {122}},\ \bibinfo {pages} {112001} (\bibinfo {year} {2019})},\
  \Eprint {http://arxiv.org/abs/1811.00559} {arXiv:1811.00559 [hep-ph]}
  \BibitemShut {NoStop}%
\bibitem [{\citenamefont {Croon}\ \emph {et~al.}(2020)\citenamefont {Croon},
  \citenamefont {Howard}, \citenamefont {Ipek},\ and\ \citenamefont
  {Tait}}]{Croon:2019ugf}%
  \BibitemOpen
  \bibfield  {author} {\bibinfo {author} {\bibfnamefont {D.}~\bibnamefont
  {Croon}}, \bibinfo {author} {\bibfnamefont {J.~N.}\ \bibnamefont {Howard}},
  \bibinfo {author} {\bibfnamefont {S.}~\bibnamefont {Ipek}}, \ and\ \bibinfo
  {author} {\bibfnamefont {T.~M.~P.}\ \bibnamefont {Tait}},\ }\href {\doibase
  10.1103/PhysRevD.101.055042} {\bibfield  {journal} {\bibinfo  {journal}
  {Phys. Rev. D}\ }\textbf {\bibinfo {volume} {101}},\ \bibinfo {pages}
  {055042} (\bibinfo {year} {2020})},\ \Eprint
  {http://arxiv.org/abs/1911.01432} {arXiv:1911.01432 [hep-ph]} \BibitemShut
  {NoStop}%
\bibitem [{\citenamefont {Berger}\ \emph {et~al.}(2020)\citenamefont {Berger},
  \citenamefont {Ipek}, \citenamefont {Tait},\ and\ \citenamefont
  {Waterbury}}]{Berger:2020maa}%
  \BibitemOpen
  \bibfield  {author} {\bibinfo {author} {\bibfnamefont {D.}~\bibnamefont
  {Berger}}, \bibinfo {author} {\bibfnamefont {S.}~\bibnamefont {Ipek}},
  \bibinfo {author} {\bibfnamefont {T.~M.~P.}\ \bibnamefont {Tait}}, \ and\
  \bibinfo {author} {\bibfnamefont {M.}~\bibnamefont {Waterbury}},\ }\href
  {\doibase 10.1007/JHEP07(2020)192} {\bibfield  {journal} {\bibinfo  {journal}
  {JHEP}\ }\textbf {\bibinfo {volume} {07}},\ \bibinfo {pages} {192} (\bibinfo
  {year} {2020})},\ \Eprint {http://arxiv.org/abs/2004.06727} {arXiv:2004.06727
  [hep-ph]} \BibitemShut {NoStop}%
\bibitem [{\citenamefont {von Harling}\ and\ \citenamefont
  {Servant}(2018)}]{vonHarling:2017yew}%
  \BibitemOpen
  \bibfield  {author} {\bibinfo {author} {\bibfnamefont {B.}~\bibnamefont {von
  Harling}}\ and\ \bibinfo {author} {\bibfnamefont {G.}~\bibnamefont
  {Servant}},\ }\href {\doibase 10.1007/JHEP01(2018)159} {\bibfield  {journal}
  {\bibinfo  {journal} {JHEP}\ }\textbf {\bibinfo {volume} {01}},\ \bibinfo
  {pages} {159} (\bibinfo {year} {2018})},\ \Eprint
  {http://arxiv.org/abs/1711.11554} {arXiv:1711.11554 [hep-ph]} \BibitemShut
  {NoStop}%
\bibitem [{\citenamefont {Aad}\ \emph {et~al.}(2020)\citenamefont {Aad} \emph
  {et~al.}}]{ATLAS:2019nkf}%
  \BibitemOpen
  \bibfield  {author} {\bibinfo {author} {\bibfnamefont {G.}~\bibnamefont
  {Aad}} \emph {et~al.} (\bibinfo {collaboration} {ATLAS}),\ }\href {\doibase
  10.1103/PhysRevD.101.012002} {\bibfield  {journal} {\bibinfo  {journal}
  {Phys. Rev. D}\ }\textbf {\bibinfo {volume} {101}},\ \bibinfo {pages}
  {012002} (\bibinfo {year} {2020})},\ \Eprint
  {http://arxiv.org/abs/1909.02845} {arXiv:1909.02845 [hep-ex]} \BibitemShut
  {NoStop}%
\bibitem [{\citenamefont {Niikura}\ \emph
  {et~al.}(2019{\natexlab{a}})\citenamefont {Niikura} \emph
  {et~al.}}]{Niikura:2017zjd}%
  \BibitemOpen
  \bibfield  {author} {\bibinfo {author} {\bibfnamefont {H.}~\bibnamefont
  {Niikura}} \emph {et~al.},\ }\href {\doibase 10.1038/s41550-019-0723-1}
  {\bibfield  {journal} {\bibinfo  {journal} {Nat. Astron.}\ }\textbf {\bibinfo
  {volume} {3}},\ \bibinfo {pages} {524} (\bibinfo {year}
  {2019}{\natexlab{a}})},\ \Eprint {http://arxiv.org/abs/1701.02151}
  {arXiv:1701.02151 [astro-ph.CO]} \BibitemShut {NoStop}%
\bibitem [{\citenamefont {Smyth}\ \emph {et~al.}(2019)\citenamefont {Smyth},
  \citenamefont {Profumo}, \citenamefont {English}, \citenamefont {Jeltema},
  \citenamefont {McKinnon},\ and\ \citenamefont
  {Guhathakurta}}]{Smyth:2019whb}%
  \BibitemOpen
  \bibfield  {author} {\bibinfo {author} {\bibfnamefont {N.}~\bibnamefont
  {Smyth}}, \bibinfo {author} {\bibfnamefont {S.}~\bibnamefont {Profumo}},
  \bibinfo {author} {\bibfnamefont {S.}~\bibnamefont {English}}, \bibinfo
  {author} {\bibfnamefont {T.}~\bibnamefont {Jeltema}}, \bibinfo {author}
  {\bibfnamefont {K.}~\bibnamefont {McKinnon}}, \ and\ \bibinfo {author}
  {\bibfnamefont {P.}~\bibnamefont {Guhathakurta}},\ }\href@noop {} {\
  (\bibinfo {year} {2019})},\ \Eprint {http://arxiv.org/abs/1910.01285}
  {arXiv:1910.01285 [astro-ph.CO]} \BibitemShut {NoStop}%
\bibitem [{\citenamefont {Allsman}\ \emph {et~al.}(2001)\citenamefont {Allsman}
  \emph {et~al.}}]{Macho:2000nvd}%
  \BibitemOpen
  \bibfield  {author} {\bibinfo {author} {\bibfnamefont {R.~A.}\ \bibnamefont
  {Allsman}} \emph {et~al.} (\bibinfo {collaboration} {Macho}),\ }\href
  {\doibase 10.1086/319636} {\bibfield  {journal} {\bibinfo  {journal}
  {Astrophys. J. Lett.}\ }\textbf {\bibinfo {volume} {550}},\ \bibinfo {pages}
  {L169} (\bibinfo {year} {2001})},\ \Eprint
  {http://arxiv.org/abs/astro-ph/0011506} {arXiv:astro-ph/0011506} \BibitemShut
  {NoStop}%
\bibitem [{\citenamefont {Alcock}\ \emph {et~al.}(2001)\citenamefont {Alcock}
  \emph {et~al.}}]{MACHO:2000bzs}%
  \BibitemOpen
  \bibfield  {author} {\bibinfo {author} {\bibfnamefont {C.}~\bibnamefont
  {Alcock}} \emph {et~al.} (\bibinfo {collaboration} {MACHO}),\ }\href
  {\doibase 10.1086/322529} {\bibfield  {journal} {\bibinfo  {journal}
  {Astrophys. J. Suppl.}\ }\textbf {\bibinfo {volume} {136}},\ \bibinfo {pages}
  {439} (\bibinfo {year} {2001})},\ \Eprint
  {http://arxiv.org/abs/astro-ph/0003392} {arXiv:astro-ph/0003392} \BibitemShut
  {NoStop}%
\bibitem [{\citenamefont {Niikura}\ \emph
  {et~al.}(2019{\natexlab{b}})\citenamefont {Niikura}, \citenamefont {Takada},
  \citenamefont {Yokoyama}, \citenamefont {Sumi},\ and\ \citenamefont
  {Masaki}}]{Niikura:2019kqi}%
  \BibitemOpen
  \bibfield  {author} {\bibinfo {author} {\bibfnamefont {H.}~\bibnamefont
  {Niikura}}, \bibinfo {author} {\bibfnamefont {M.}~\bibnamefont {Takada}},
  \bibinfo {author} {\bibfnamefont {S.}~\bibnamefont {Yokoyama}}, \bibinfo
  {author} {\bibfnamefont {T.}~\bibnamefont {Sumi}}, \ and\ \bibinfo {author}
  {\bibfnamefont {S.}~\bibnamefont {Masaki}},\ }\href {\doibase
  10.1103/PhysRevD.99.083503} {\bibfield  {journal} {\bibinfo  {journal} {Phys.
  Rev.}\ }\textbf {\bibinfo {volume} {D99}},\ \bibinfo {pages} {083503}
  (\bibinfo {year} {2019}{\natexlab{b}})},\ \Eprint
  {http://arxiv.org/abs/1901.07120} {arXiv:1901.07120 [astro-ph.CO]}
  \BibitemShut {NoStop}%
\bibitem [{\citenamefont {Oguri}\ \emph {et~al.}(2018)\citenamefont {Oguri},
  \citenamefont {Diego}, \citenamefont {Kaiser}, \citenamefont {Kelly},\ and\
  \citenamefont {Broadhurst}}]{Oguri:2017ock}%
  \BibitemOpen
  \bibfield  {author} {\bibinfo {author} {\bibfnamefont {M.}~\bibnamefont
  {Oguri}}, \bibinfo {author} {\bibfnamefont {J.~M.}\ \bibnamefont {Diego}},
  \bibinfo {author} {\bibfnamefont {N.}~\bibnamefont {Kaiser}}, \bibinfo
  {author} {\bibfnamefont {P.~L.}\ \bibnamefont {Kelly}}, \ and\ \bibinfo
  {author} {\bibfnamefont {T.}~\bibnamefont {Broadhurst}},\ }\href {\doibase
  10.1103/PhysRevD.97.023518} {\bibfield  {journal} {\bibinfo  {journal} {Phys.
  Rev. D}\ }\textbf {\bibinfo {volume} {97}},\ \bibinfo {pages} {023518}
  (\bibinfo {year} {2018})},\ \Eprint {http://arxiv.org/abs/1710.00148}
  {arXiv:1710.00148 [astro-ph.CO]} \BibitemShut {NoStop}%
\bibitem [{\citenamefont {Griest}\ \emph {et~al.}(2014)\citenamefont {Griest},
  \citenamefont {Cieplak},\ and\ \citenamefont {Lehner}}]{Griest:2013aaa}%
  \BibitemOpen
  \bibfield  {author} {\bibinfo {author} {\bibfnamefont {K.}~\bibnamefont
  {Griest}}, \bibinfo {author} {\bibfnamefont {A.~M.}\ \bibnamefont {Cieplak}},
  \ and\ \bibinfo {author} {\bibfnamefont {M.~J.}\ \bibnamefont {Lehner}},\
  }\href {\doibase 10.1088/0004-637X/786/2/158} {\bibfield  {journal} {\bibinfo
   {journal} {Astrophys. J.}\ }\textbf {\bibinfo {volume} {786}},\ \bibinfo
  {pages} {158} (\bibinfo {year} {2014})},\ \Eprint
  {http://arxiv.org/abs/1307.5798} {arXiv:1307.5798 [astro-ph.CO]} \BibitemShut
  {NoStop}%
\bibitem [{\citenamefont {Lentati}\ \emph {et~al.}(2015)\citenamefont {Lentati}
  \emph {et~al.}}]{Lentati:2015qwp}%
  \BibitemOpen
  \bibfield  {author} {\bibinfo {author} {\bibfnamefont {L.}~\bibnamefont
  {Lentati}} \emph {et~al.},\ }\href {\doibase 10.1093/mnras/stv1538}
  {\bibfield  {journal} {\bibinfo  {journal} {Mon. Not. Roy. Astron. Soc.}\
  }\textbf {\bibinfo {volume} {453}},\ \bibinfo {pages} {2576} (\bibinfo {year}
  {2015})},\ \Eprint {http://arxiv.org/abs/1504.03692} {arXiv:1504.03692
  [astro-ph.CO]} \BibitemShut {NoStop}%
\bibitem [{\citenamefont {Shannon}\ \emph {et~al.}(2015)\citenamefont {Shannon}
  \emph {et~al.}}]{Shannon:2015ect}%
  \BibitemOpen
  \bibfield  {author} {\bibinfo {author} {\bibfnamefont {R.}~\bibnamefont
  {Shannon}} \emph {et~al.},\ }\href {\doibase 10.1126/science.aab1910}
  {\bibfield  {journal} {\bibinfo  {journal} {Science}\ }\textbf {\bibinfo
  {volume} {349}},\ \bibinfo {pages} {1522} (\bibinfo {year} {2015})},\ \Eprint
  {http://arxiv.org/abs/1509.07320} {arXiv:1509.07320 [astro-ph.CO]}
  \BibitemShut {NoStop}%
\bibitem [{\citenamefont {Arzoumanian}\ \emph {et~al.}(2018)\citenamefont
  {Arzoumanian} \emph {et~al.}}]{NANOGRAV:2018hou}%
  \BibitemOpen
  \bibfield  {author} {\bibinfo {author} {\bibfnamefont {Z.}~\bibnamefont
  {Arzoumanian}} \emph {et~al.} (\bibinfo {collaboration} {NANOGRAV}),\ }\href
  {\doibase 10.3847/1538-4357/aabd3b} {\bibfield  {journal} {\bibinfo
  {journal} {Astrophys. J.}\ }\textbf {\bibinfo {volume} {859}},\ \bibinfo
  {pages} {47} (\bibinfo {year} {2018})},\ \Eprint
  {http://arxiv.org/abs/1801.02617} {arXiv:1801.02617 [astro-ph.HE]}
  \BibitemShut {NoStop}%
\bibitem [{\citenamefont {Aggarwal}\ \emph {et~al.}(2019)\citenamefont
  {Aggarwal} \emph {et~al.}}]{Aggarwal:2018mgp}%
  \BibitemOpen
  \bibfield  {author} {\bibinfo {author} {\bibfnamefont {K.}~\bibnamefont
  {Aggarwal}} \emph {et~al.},\ }\href {\doibase 10.3847/1538-4357/ab2236}
  {\bibfield  {journal} {\bibinfo  {journal} {Astrophys. J.}\ }\textbf
  {\bibinfo {volume} {880}},\ \bibinfo {pages} {2} (\bibinfo {year} {2019})},\
  \Eprint {http://arxiv.org/abs/1812.11585} {arXiv:1812.11585 [astro-ph.GA]}
  \BibitemShut {NoStop}%
\bibitem [{\citenamefont {Zhao}\ \emph {et~al.}(2013)\citenamefont {Zhao},
  \citenamefont {Zhang}, \citenamefont {You},\ and\ \citenamefont
  {Zhu}}]{Zhao:2013bba}%
  \BibitemOpen
  \bibfield  {author} {\bibinfo {author} {\bibfnamefont {W.}~\bibnamefont
  {Zhao}}, \bibinfo {author} {\bibfnamefont {Y.}~\bibnamefont {Zhang}},
  \bibinfo {author} {\bibfnamefont {X.-P.}\ \bibnamefont {You}}, \ and\
  \bibinfo {author} {\bibfnamefont {Z.-H.}\ \bibnamefont {Zhu}},\ }\href
  {\doibase 10.1103/PhysRevD.87.124012} {\bibfield  {journal} {\bibinfo
  {journal} {Phys. Rev. D}\ }\textbf {\bibinfo {volume} {87}},\ \bibinfo
  {pages} {124012} (\bibinfo {year} {2013})},\ \Eprint
  {http://arxiv.org/abs/1303.6718} {arXiv:1303.6718 [astro-ph.CO]} \BibitemShut
  {NoStop}%
\bibitem [{\citenamefont {Amaro-Seoane}\ \emph {et~al.}(2017)\citenamefont
  {Amaro-Seoane} \emph {et~al.}}]{LISA:2017}%
  \BibitemOpen
  \bibfield  {author} {\bibinfo {author} {\bibfnamefont {P.}~\bibnamefont
  {Amaro-Seoane}} \emph {et~al.},\ }\href@noop {} {\  (\bibinfo {year}
  {2017})},\ \Eprint {http://arxiv.org/abs/1702.00786} {arXiv:1702.00786
  [astro-ph.IM]} \BibitemShut {NoStop}%
\bibitem [{\citenamefont {Yagi}\ and\ \citenamefont
  {Seto}(2011)}]{Yagi:2011wg}%
  \BibitemOpen
  \bibfield  {author} {\bibinfo {author} {\bibfnamefont {K.}~\bibnamefont
  {Yagi}}\ and\ \bibinfo {author} {\bibfnamefont {N.}~\bibnamefont {Seto}},\
  }\href {\doibase 10.1103/PhysRevD.83.044011} {\bibfield  {journal} {\bibinfo
  {journal} {Phys. Rev. D}\ }\textbf {\bibinfo {volume} {83}},\ \bibinfo
  {pages} {044011} (\bibinfo {year} {2011})},\ \bibinfo {note} {[Erratum:
  Phys.Rev.D 95, 109901 (2017)]},\ \Eprint {http://arxiv.org/abs/1101.3940}
  {arXiv:1101.3940 [astro-ph.CO]} \BibitemShut {NoStop}%
\bibitem [{\citenamefont {Abbott}\ \emph
  {et~al.}(2017{\natexlab{a}})\citenamefont {Abbott} \emph
  {et~al.}}]{LIGOScientific:2016wof}%
  \BibitemOpen
  \bibfield  {author} {\bibinfo {author} {\bibfnamefont {B.~P.}\ \bibnamefont
  {Abbott}} \emph {et~al.} (\bibinfo {collaboration} {LIGO Scientific}),\
  }\href {\doibase 10.1088/1361-6382/aa51f4} {\bibfield  {journal} {\bibinfo
  {journal} {Class. Quant. Grav.}\ }\textbf {\bibinfo {volume} {34}},\ \bibinfo
  {pages} {044001} (\bibinfo {year} {2017}{\natexlab{a}})},\ \Eprint
  {http://arxiv.org/abs/1607.08697} {arXiv:1607.08697 [astro-ph.IM]}
  \BibitemShut {NoStop}%
\bibitem [{\citenamefont {Moore}\ \emph {et~al.}(2015)\citenamefont {Moore},
  \citenamefont {Cole},\ and\ \citenamefont {Berry}}]{Moore:2014lga}%
  \BibitemOpen
  \bibfield  {author} {\bibinfo {author} {\bibfnamefont {C.~J.}\ \bibnamefont
  {Moore}}, \bibinfo {author} {\bibfnamefont {R.~H.}\ \bibnamefont {Cole}}, \
  and\ \bibinfo {author} {\bibfnamefont {C.~P.~L.}\ \bibnamefont {Berry}},\
  }\href {\doibase 10.1088/0264-9381/32/1/015014} {\bibfield  {journal}
  {\bibinfo  {journal} {Class. Quant. Grav.}\ }\textbf {\bibinfo {volume}
  {32}},\ \bibinfo {pages} {015014} (\bibinfo {year} {2015})},\ \Eprint
  {http://arxiv.org/abs/1408.0740} {arXiv:1408.0740 [gr-qc]} \BibitemShut
  {NoStop}%
\bibitem [{\citenamefont {Sathyaprakash}\ and\ \citenamefont
  {Schutz}(2009)}]{Sathyaprakash:2009xs}%
  \BibitemOpen
  \bibfield  {author} {\bibinfo {author} {\bibfnamefont {B.~S.}\ \bibnamefont
  {Sathyaprakash}}\ and\ \bibinfo {author} {\bibfnamefont {B.~F.}\ \bibnamefont
  {Schutz}},\ }\href {\doibase 10.12942/lrr-2009-2} {\bibfield  {journal}
  {\bibinfo  {journal} {Living Rev. Rel.}\ }\textbf {\bibinfo {volume} {12}},\
  \bibinfo {pages} {2} (\bibinfo {year} {2009})},\ \Eprint
  {http://arxiv.org/abs/0903.0338} {arXiv:0903.0338 [gr-qc]} \BibitemShut
  {NoStop}%
\bibitem [{\citenamefont {Sesana}\ \emph {et~al.}(2021)\citenamefont {Sesana}
  \emph {et~al.}}]{Sesana:2019vho}%
  \BibitemOpen
  \bibfield  {author} {\bibinfo {author} {\bibfnamefont {A.}~\bibnamefont
  {Sesana}} \emph {et~al.},\ }\href {\doibase 10.1007/s10686-021-09709-9}
  {\bibfield  {journal} {\bibinfo  {journal} {Exper. Astron.}\ }\textbf
  {\bibinfo {volume} {51}},\ \bibinfo {pages} {1333} (\bibinfo {year}
  {2021})},\ \Eprint {http://arxiv.org/abs/1908.11391} {arXiv:1908.11391
  [astro-ph.IM]} \BibitemShut {NoStop}%
\bibitem [{\citenamefont {Coleman}(2019)}]{Coleman:2018ozp}%
  \BibitemOpen
  \bibfield  {author} {\bibinfo {author} {\bibfnamefont {J.}~\bibnamefont
  {Coleman}} (\bibinfo {collaboration} {MAGIS-100}),\ }\href {\doibase
  10.22323/1.340.0021} {\bibfield  {journal} {\bibinfo  {journal} {PoS}\
  }\textbf {\bibinfo {volume} {ICHEP2018}},\ \bibinfo {pages} {021} (\bibinfo
  {year} {2019})},\ \Eprint {http://arxiv.org/abs/1812.00482} {arXiv:1812.00482
  [physics.ins-det]} \BibitemShut {NoStop}%
\bibitem [{\citenamefont {El-Neaj}\ \emph {et~al.}(2020)\citenamefont {El-Neaj}
  \emph {et~al.}}]{AEDGE:2019nxb}%
  \BibitemOpen
  \bibfield  {author} {\bibinfo {author} {\bibfnamefont {Y.~A.}\ \bibnamefont
  {El-Neaj}} \emph {et~al.} (\bibinfo {collaboration} {AEDGE}),\ }\href
  {\doibase 10.1140/epjqt/s40507-020-0080-0} {\bibfield  {journal} {\bibinfo
  {journal} {EPJ Quant. Technol.}\ }\textbf {\bibinfo {volume} {7}},\ \bibinfo
  {pages} {6} (\bibinfo {year} {2020})},\ \Eprint
  {http://arxiv.org/abs/1908.00802} {arXiv:1908.00802 [gr-qc]} \BibitemShut
  {NoStop}%
\bibitem [{\citenamefont {Badurina}\ \emph {et~al.}(2020)\citenamefont
  {Badurina} \emph {et~al.}}]{Badurina:2019hst}%
  \BibitemOpen
  \bibfield  {author} {\bibinfo {author} {\bibfnamefont {L.}~\bibnamefont
  {Badurina}} \emph {et~al.},\ }\href {\doibase 10.1088/1475-7516/2020/05/011}
  {\bibfield  {journal} {\bibinfo  {journal} {JCAP}\ }\textbf {\bibinfo
  {volume} {05}},\ \bibinfo {pages} {011} (\bibinfo {year} {2020})},\ \Eprint
  {http://arxiv.org/abs/1911.11755} {arXiv:1911.11755 [astro-ph.CO]}
  \BibitemShut {NoStop}%
\bibitem [{\citenamefont {Abbott}\ \emph
  {et~al.}(2017{\natexlab{b}})\citenamefont {Abbott} \emph
  {et~al.}}]{LIGOScientific:2016jlg}%
  \BibitemOpen
  \bibfield  {author} {\bibinfo {author} {\bibfnamefont {B.~P.}\ \bibnamefont
  {Abbott}} \emph {et~al.} (\bibinfo {collaboration} {LIGO Scientific,
  Virgo}),\ }\href {\doibase 10.1103/PhysRevLett.118.121101} {\bibfield
  {journal} {\bibinfo  {journal} {Phys. Rev. Lett.}\ }\textbf {\bibinfo
  {volume} {118}},\ \bibinfo {pages} {121101} (\bibinfo {year}
  {2017}{\natexlab{b}})},\ \bibinfo {note} {[Erratum: Phys.Rev.Lett. 119,
  029901 (2017)]},\ \Eprint {http://arxiv.org/abs/1612.02029} {arXiv:1612.02029
  [gr-qc]} \BibitemShut {NoStop}%
\bibitem [{\citenamefont {Fukushima}(2004)}]{Fukushima:2003fw}%
  \BibitemOpen
  \bibfield  {author} {\bibinfo {author} {\bibfnamefont {K.}~\bibnamefont
  {Fukushima}},\ }\href {\doibase 10.1016/j.physletb.2004.04.027} {\bibfield
  {journal} {\bibinfo  {journal} {Phys. Lett. B}\ }\textbf {\bibinfo {volume}
  {591}},\ \bibinfo {pages} {277} (\bibinfo {year} {2004})},\ \Eprint
  {http://arxiv.org/abs/hep-ph/0310121} {arXiv:hep-ph/0310121} \BibitemShut
  {NoStop}%
\bibitem [{\citenamefont {Ratti}\ \emph {et~al.}(2006)\citenamefont {Ratti},
  \citenamefont {Thaler},\ and\ \citenamefont {Weise}}]{Ratti:2005jh}%
  \BibitemOpen
  \bibfield  {author} {\bibinfo {author} {\bibfnamefont {C.}~\bibnamefont
  {Ratti}}, \bibinfo {author} {\bibfnamefont {M.~A.}\ \bibnamefont {Thaler}}, \
  and\ \bibinfo {author} {\bibfnamefont {W.}~\bibnamefont {Weise}},\ }\href
  {\doibase 10.1103/PhysRevD.73.014019} {\bibfield  {journal} {\bibinfo
  {journal} {Phys. Rev. D}\ }\textbf {\bibinfo {volume} {73}},\ \bibinfo
  {pages} {014019} (\bibinfo {year} {2006})},\ \Eprint
  {http://arxiv.org/abs/hep-ph/0506234} {arXiv:hep-ph/0506234} \BibitemShut
  {NoStop}%
\bibitem [{\citenamefont {Helmboldt}\ \emph {et~al.}(2019)\citenamefont
  {Helmboldt}, \citenamefont {Kubo},\ and\ \citenamefont {van~der
  Woude}}]{Helmboldt:2019pan}%
  \BibitemOpen
  \bibfield  {author} {\bibinfo {author} {\bibfnamefont {A.~J.}\ \bibnamefont
  {Helmboldt}}, \bibinfo {author} {\bibfnamefont {J.}~\bibnamefont {Kubo}}, \
  and\ \bibinfo {author} {\bibfnamefont {S.}~\bibnamefont {van~der Woude}},\
  }\href {\doibase 10.1103/PhysRevD.100.055025} {\bibfield  {journal} {\bibinfo
   {journal} {Phys. Rev. D}\ }\textbf {\bibinfo {volume} {100}},\ \bibinfo
  {pages} {055025} (\bibinfo {year} {2019})},\ \Eprint
  {http://arxiv.org/abs/1904.07891} {arXiv:1904.07891 [hep-ph]} \BibitemShut
  {NoStop}%
\bibitem [{\citenamefont {Wands}(1999)}]{Wands:1998yp}%
  \BibitemOpen
  \bibfield  {author} {\bibinfo {author} {\bibfnamefont {D.}~\bibnamefont
  {Wands}},\ }\href {\doibase 10.1103/PhysRevD.60.023507} {\bibfield  {journal}
  {\bibinfo  {journal} {Phys. Rev. D}\ }\textbf {\bibinfo {volume} {60}},\
  \bibinfo {pages} {023507} (\bibinfo {year} {1999})},\ \Eprint
  {http://arxiv.org/abs/gr-qc/9809062} {arXiv:gr-qc/9809062} \BibitemShut
  {NoStop}%
\bibitem [{\citenamefont {Leach}\ and\ \citenamefont
  {Liddle}(2001)}]{Leach:2000yw}%
  \BibitemOpen
  \bibfield  {author} {\bibinfo {author} {\bibfnamefont {S.~M.}\ \bibnamefont
  {Leach}}\ and\ \bibinfo {author} {\bibfnamefont {A.~R.}\ \bibnamefont
  {Liddle}},\ }\href {\doibase 10.1103/PhysRevD.63.043508} {\bibfield
  {journal} {\bibinfo  {journal} {Phys. Rev. D}\ }\textbf {\bibinfo {volume}
  {63}},\ \bibinfo {pages} {043508} (\bibinfo {year} {2001})},\ \Eprint
  {http://arxiv.org/abs/astro-ph/0010082} {arXiv:astro-ph/0010082} \BibitemShut
  {NoStop}%
\bibitem [{\citenamefont {Leach}\ \emph {et~al.}(2001)\citenamefont {Leach},
  \citenamefont {Sasaki}, \citenamefont {Wands},\ and\ \citenamefont
  {Liddle}}]{Leach:2001zf}%
  \BibitemOpen
  \bibfield  {author} {\bibinfo {author} {\bibfnamefont {S.~M.}\ \bibnamefont
  {Leach}}, \bibinfo {author} {\bibfnamefont {M.}~\bibnamefont {Sasaki}},
  \bibinfo {author} {\bibfnamefont {D.}~\bibnamefont {Wands}}, \ and\ \bibinfo
  {author} {\bibfnamefont {A.~R.}\ \bibnamefont {Liddle}},\ }\href {\doibase
  10.1103/PhysRevD.64.023512} {\bibfield  {journal} {\bibinfo  {journal} {Phys.
  Rev. D}\ }\textbf {\bibinfo {volume} {64}},\ \bibinfo {pages} {023512}
  (\bibinfo {year} {2001})},\ \Eprint {http://arxiv.org/abs/astro-ph/0101406}
  {arXiv:astro-ph/0101406} \BibitemShut {NoStop}%
\bibitem [{\citenamefont {Biagetti}\ \emph {et~al.}(2018)\citenamefont
  {Biagetti}, \citenamefont {Franciolini}, \citenamefont {Kehagias},\ and\
  \citenamefont {Riotto}}]{Biagetti:2018pjj}%
  \BibitemOpen
  \bibfield  {author} {\bibinfo {author} {\bibfnamefont {M.}~\bibnamefont
  {Biagetti}}, \bibinfo {author} {\bibfnamefont {G.}~\bibnamefont
  {Franciolini}}, \bibinfo {author} {\bibfnamefont {A.}~\bibnamefont
  {Kehagias}}, \ and\ \bibinfo {author} {\bibfnamefont {A.}~\bibnamefont
  {Riotto}},\ }\href {\doibase 10.1088/1475-7516/2018/07/032} {\bibfield
  {journal} {\bibinfo  {journal} {JCAP}\ }\textbf {\bibinfo {volume} {07}},\
  \bibinfo {pages} {032} (\bibinfo {year} {2018})},\ \Eprint
  {http://arxiv.org/abs/1804.07124} {arXiv:1804.07124 [astro-ph.CO]}
  \BibitemShut {NoStop}%
\bibitem [{\citenamefont {Musco}\ and\ \citenamefont
  {Miller}(2013)}]{Musco:2012au}%
  \BibitemOpen
  \bibfield  {author} {\bibinfo {author} {\bibfnamefont {I.}~\bibnamefont
  {Musco}}\ and\ \bibinfo {author} {\bibfnamefont {J.~C.}\ \bibnamefont
  {Miller}},\ }\href {\doibase 10.1088/0264-9381/30/14/145009} {\bibfield
  {journal} {\bibinfo  {journal} {Class. Quant. Grav.}\ }\textbf {\bibinfo
  {volume} {30}},\ \bibinfo {pages} {145009} (\bibinfo {year} {2013})},\
  \Eprint {http://arxiv.org/abs/1201.2379} {arXiv:1201.2379 [gr-qc]}
  \BibitemShut {NoStop}%
\bibitem [{\citenamefont {Gow}\ \emph {et~al.}(2021)\citenamefont {Gow},
  \citenamefont {Byrnes}, \citenamefont {Cole},\ and\ \citenamefont
  {Young}}]{Gow:2020bzo}%
  \BibitemOpen
  \bibfield  {author} {\bibinfo {author} {\bibfnamefont {A.~D.}\ \bibnamefont
  {Gow}}, \bibinfo {author} {\bibfnamefont {C.~T.}\ \bibnamefont {Byrnes}},
  \bibinfo {author} {\bibfnamefont {P.~S.}\ \bibnamefont {Cole}}, \ and\
  \bibinfo {author} {\bibfnamefont {S.}~\bibnamefont {Young}},\ }\href
  {\doibase 10.1088/1475-7516/2021/02/002} {\bibfield  {journal} {\bibinfo
  {journal} {JCAP}\ }\textbf {\bibinfo {volume} {02}},\ \bibinfo {pages} {002}
  (\bibinfo {year} {2021})},\ \Eprint {http://arxiv.org/abs/2008.03289}
  {arXiv:2008.03289 [astro-ph.CO]} \BibitemShut {NoStop}%
\bibitem [{\citenamefont {De~Luca}\ \emph
  {et~al.}(2020{\natexlab{a}})\citenamefont {De~Luca}, \citenamefont
  {Franciolini},\ and\ \citenamefont {Riotto}}]{DeLuca:2020ioi}%
  \BibitemOpen
  \bibfield  {author} {\bibinfo {author} {\bibfnamefont {V.}~\bibnamefont
  {De~Luca}}, \bibinfo {author} {\bibfnamefont {G.}~\bibnamefont
  {Franciolini}}, \ and\ \bibinfo {author} {\bibfnamefont {A.}~\bibnamefont
  {Riotto}},\ }\href {\doibase 10.1016/j.physletb.2020.135550} {\bibfield
  {journal} {\bibinfo  {journal} {Phys. Lett. B}\ }\textbf {\bibinfo {volume}
  {807}},\ \bibinfo {pages} {135550} (\bibinfo {year} {2020}{\natexlab{a}})},\
  \Eprint {http://arxiv.org/abs/2001.04371} {arXiv:2001.04371 [astro-ph.CO]}
  \BibitemShut {NoStop}%
\bibitem [{sup()}]{supmat}%
  \BibitemOpen
  \href@noop {} {}\bibinfo {howpublished} {See Supplemental Material [url] for
  details of PNJL model for a high temperature QCD transition and computation
  of theory parameters, which includes
  Refs.~\cite{Ratti:2005jh,Helmboldt:2019pan,Reichert:2021cvs,Borsanyi:2016ksw}.}\BibitemShut
  {Stop}%
\bibitem [{\citenamefont {Takhistov}\ \emph {et~al.}(2021)\citenamefont
  {Takhistov}, \citenamefont {Fuller},\ and\ \citenamefont
  {Kusenko}}]{Takhistov:2020vxs}%
  \BibitemOpen
  \bibfield  {author} {\bibinfo {author} {\bibfnamefont {V.}~\bibnamefont
  {Takhistov}}, \bibinfo {author} {\bibfnamefont {G.~M.}\ \bibnamefont
  {Fuller}}, \ and\ \bibinfo {author} {\bibfnamefont {A.}~\bibnamefont
  {Kusenko}},\ }\href {\doibase 10.1103/PhysRevLett.126.071101} {\bibfield
  {journal} {\bibinfo  {journal} {Phys. Rev. Lett.}\ }\textbf {\bibinfo
  {volume} {126}},\ \bibinfo {pages} {071101} (\bibinfo {year} {2021})},\
  \Eprint {http://arxiv.org/abs/2008.12780} {arXiv:2008.12780 [astro-ph.HE]}
  \BibitemShut {NoStop}%
\bibitem [{\citenamefont {G\'enolini}\ \emph {et~al.}(2020)\citenamefont
  {G\'enolini}, \citenamefont {Serpico},\ and\ \citenamefont
  {Tinyakov}}]{Genolini:2020ejw}%
  \BibitemOpen
  \bibfield  {author} {\bibinfo {author} {\bibfnamefont {Y.}~\bibnamefont
  {G\'enolini}}, \bibinfo {author} {\bibfnamefont {P.}~\bibnamefont {Serpico}},
  \ and\ \bibinfo {author} {\bibfnamefont {P.}~\bibnamefont {Tinyakov}},\
  }\href {\doibase 10.1103/PhysRevD.102.083004} {\bibfield  {journal} {\bibinfo
   {journal} {Phys. Rev. D}\ }\textbf {\bibinfo {volume} {102}},\ \bibinfo
  {pages} {083004} (\bibinfo {year} {2020})},\ \Eprint
  {http://arxiv.org/abs/2006.16975} {arXiv:2006.16975 [astro-ph.HE]}
  \BibitemShut {NoStop}%
\bibitem [{\citenamefont {Dasgupta}\ \emph {et~al.}(2021)\citenamefont
  {Dasgupta}, \citenamefont {Laha},\ and\ \citenamefont
  {Ray}}]{Dasgupta:2020mqg}%
  \BibitemOpen
  \bibfield  {author} {\bibinfo {author} {\bibfnamefont {B.}~\bibnamefont
  {Dasgupta}}, \bibinfo {author} {\bibfnamefont {R.}~\bibnamefont {Laha}}, \
  and\ \bibinfo {author} {\bibfnamefont {A.}~\bibnamefont {Ray}},\ }\href
  {\doibase 10.1103/PhysRevLett.126.141105} {\bibfield  {journal} {\bibinfo
  {journal} {Phys. Rev. Lett.}\ }\textbf {\bibinfo {volume} {126}},\ \bibinfo
  {pages} {141105} (\bibinfo {year} {2021})},\ \Eprint
  {http://arxiv.org/abs/2009.01825} {arXiv:2009.01825 [astro-ph.HE]}
  \BibitemShut {NoStop}%
\bibitem [{\citenamefont {Sugiyama}\ \emph {et~al.}(2019)\citenamefont
  {Sugiyama}, \citenamefont {Kurita},\ and\ \citenamefont
  {Takada}}]{Sugiyama:2019dgt}%
  \BibitemOpen
  \bibfield  {author} {\bibinfo {author} {\bibfnamefont {S.}~\bibnamefont
  {Sugiyama}}, \bibinfo {author} {\bibfnamefont {T.}~\bibnamefont {Kurita}}, \
  and\ \bibinfo {author} {\bibfnamefont {M.}~\bibnamefont {Takada}},\
  }\href@noop {} {\  (\bibinfo {year} {2019})},\ \Eprint
  {http://arxiv.org/abs/1905.06066} {arXiv:1905.06066 [astro-ph.CO]}
  \BibitemShut {NoStop}%
\bibitem [{\citenamefont {Sugiyama}\ \emph {et~al.}(2021)\citenamefont
  {Sugiyama}, \citenamefont {Takhistov}, \citenamefont {Vitagliano},
  \citenamefont {Kusenko}, \citenamefont {Sasaki},\ and\ \citenamefont
  {Takada}}]{Sugiyama:2020roc}%
  \BibitemOpen
  \bibfield  {author} {\bibinfo {author} {\bibfnamefont {S.}~\bibnamefont
  {Sugiyama}}, \bibinfo {author} {\bibfnamefont {V.}~\bibnamefont {Takhistov}},
  \bibinfo {author} {\bibfnamefont {E.}~\bibnamefont {Vitagliano}}, \bibinfo
  {author} {\bibfnamefont {A.}~\bibnamefont {Kusenko}}, \bibinfo {author}
  {\bibfnamefont {M.}~\bibnamefont {Sasaki}}, \ and\ \bibinfo {author}
  {\bibfnamefont {M.}~\bibnamefont {Takada}},\ }\href {\doibase
  10.1016/j.physletb.2021.136097} {\bibfield  {journal} {\bibinfo  {journal}
  {Phys. Lett. B}\ }\textbf {\bibinfo {volume} {814}},\ \bibinfo {pages}
  {136097} (\bibinfo {year} {2021})},\ \Eprint
  {http://arxiv.org/abs/2010.02189} {arXiv:2010.02189 [astro-ph.CO]}
  \BibitemShut {NoStop}%
\bibitem [{\citenamefont {Cai}\ \emph {et~al.}(2019)\citenamefont {Cai},
  \citenamefont {Pi},\ and\ \citenamefont {Sasaki}}]{Cai:2018dig}%
  \BibitemOpen
  \bibfield  {author} {\bibinfo {author} {\bibfnamefont {R.-g.}\ \bibnamefont
  {Cai}}, \bibinfo {author} {\bibfnamefont {S.}~\bibnamefont {Pi}}, \ and\
  \bibinfo {author} {\bibfnamefont {M.}~\bibnamefont {Sasaki}},\ }\href
  {\doibase 10.1103/PhysRevLett.122.201101} {\bibfield  {journal} {\bibinfo
  {journal} {Phys. Rev. Lett.}\ }\textbf {\bibinfo {volume} {122}},\ \bibinfo
  {pages} {201101} (\bibinfo {year} {2019})},\ \Eprint
  {http://arxiv.org/abs/1810.11000} {arXiv:1810.11000 [astro-ph.CO]}
  \BibitemShut {NoStop}%
\bibitem [{\citenamefont {Ananda}\ \emph {et~al.}(2007)\citenamefont {Ananda},
  \citenamefont {Clarkson},\ and\ \citenamefont {Wands}}]{Ananda:2006af}%
  \BibitemOpen
  \bibfield  {author} {\bibinfo {author} {\bibfnamefont {K.~N.}\ \bibnamefont
  {Ananda}}, \bibinfo {author} {\bibfnamefont {C.}~\bibnamefont {Clarkson}}, \
  and\ \bibinfo {author} {\bibfnamefont {D.}~\bibnamefont {Wands}},\ }\href
  {\doibase 10.1103/PhysRevD.75.123518} {\bibfield  {journal} {\bibinfo
  {journal} {Phys. Rev. D}\ }\textbf {\bibinfo {volume} {75}},\ \bibinfo
  {pages} {123518} (\bibinfo {year} {2007})},\ \Eprint
  {http://arxiv.org/abs/gr-qc/0612013} {arXiv:gr-qc/0612013} \BibitemShut
  {NoStop}%
\bibitem [{\citenamefont {Kohri}\ and\ \citenamefont
  {Terada}(2018)}]{Kohri:2018awv}%
  \BibitemOpen
  \bibfield  {author} {\bibinfo {author} {\bibfnamefont {K.}~\bibnamefont
  {Kohri}}\ and\ \bibinfo {author} {\bibfnamefont {T.}~\bibnamefont {Terada}},\
  }\href {\doibase 10.1103/PhysRevD.97.123532} {\bibfield  {journal} {\bibinfo
  {journal} {Phys. Rev. D}\ }\textbf {\bibinfo {volume} {97}},\ \bibinfo
  {pages} {123532} (\bibinfo {year} {2018})},\ \Eprint
  {http://arxiv.org/abs/1804.08577} {arXiv:1804.08577 [gr-qc]} \BibitemShut
  {NoStop}%
\bibitem [{\citenamefont {Espinosa}\ \emph {et~al.}(2018)\citenamefont
  {Espinosa}, \citenamefont {Racco},\ and\ \citenamefont
  {Riotto}}]{Espinosa:2018eve}%
  \BibitemOpen
  \bibfield  {author} {\bibinfo {author} {\bibfnamefont {J.~R.}\ \bibnamefont
  {Espinosa}}, \bibinfo {author} {\bibfnamefont {D.}~\bibnamefont {Racco}}, \
  and\ \bibinfo {author} {\bibfnamefont {A.}~\bibnamefont {Riotto}},\ }\href
  {\doibase 10.1088/1475-7516/2018/09/012} {\bibfield  {journal} {\bibinfo
  {journal} {JCAP}\ }\textbf {\bibinfo {volume} {09}},\ \bibinfo {pages} {012}
  (\bibinfo {year} {2018})},\ \Eprint {http://arxiv.org/abs/1804.07732}
  {arXiv:1804.07732 [hep-ph]} \BibitemShut {NoStop}%
\bibitem [{\citenamefont {Dom\`enech}(2021)}]{Domenech:2021ztg}%
  \BibitemOpen
  \bibfield  {author} {\bibinfo {author} {\bibfnamefont {G.}~\bibnamefont
  {Dom\`enech}},\ }\href {\doibase 10.3390/universe7110398} {\bibfield
  {journal} {\bibinfo  {journal} {Universe}\ }\textbf {\bibinfo {volume} {7}},\
  \bibinfo {pages} {398} (\bibinfo {year} {2021})},\ \Eprint
  {http://arxiv.org/abs/2109.01398} {arXiv:2109.01398 [gr-qc]} \BibitemShut
  {NoStop}%
\bibitem [{\citenamefont {Arzoumanian}\ \emph {et~al.}(2020)\citenamefont
  {Arzoumanian} \emph {et~al.}}]{NANOGrav:2020bcs}%
  \BibitemOpen
  \bibfield  {author} {\bibinfo {author} {\bibfnamefont {Z.}~\bibnamefont
  {Arzoumanian}} \emph {et~al.} (\bibinfo {collaboration} {NANOGrav}),\ }\href
  {\doibase 10.3847/2041-8213/abd401} {\bibfield  {journal} {\bibinfo
  {journal} {Astrophys. J. Lett.}\ }\textbf {\bibinfo {volume} {905}},\
  \bibinfo {pages} {L34} (\bibinfo {year} {2020})},\ \Eprint
  {http://arxiv.org/abs/2009.04496} {arXiv:2009.04496 [astro-ph.HE]}
  \BibitemShut {NoStop}%
\bibitem [{\citenamefont {De~Luca}\ \emph
  {et~al.}(2020{\natexlab{b}})\citenamefont {De~Luca}, \citenamefont
  {Franciolini},\ and\ \citenamefont {Riotto}}]{DeLuca:2020agl}%
  \BibitemOpen
  \bibfield  {author} {\bibinfo {author} {\bibfnamefont {V.}~\bibnamefont
  {De~Luca}}, \bibinfo {author} {\bibfnamefont {G.}~\bibnamefont
  {Franciolini}}, \ and\ \bibinfo {author} {\bibfnamefont {A.}~\bibnamefont
  {Riotto}},\ }\href@noop {} {\  (\bibinfo {year} {2020}{\natexlab{b}})},\
  \Eprint {http://arxiv.org/abs/2009.08268} {arXiv:2009.08268 [astro-ph.CO]}
  \BibitemShut {NoStop}%
\bibitem [{\citenamefont {Hajkarim}\ and\ \citenamefont
  {Schaffner-Bielich}(2020)}]{Hajkarim:2019nbx}%
  \BibitemOpen
  \bibfield  {author} {\bibinfo {author} {\bibfnamefont {F.}~\bibnamefont
  {Hajkarim}}\ and\ \bibinfo {author} {\bibfnamefont {J.}~\bibnamefont
  {Schaffner-Bielich}},\ }\href {\doibase 10.1103/PhysRevD.101.043522}
  {\bibfield  {journal} {\bibinfo  {journal} {Phys. Rev. D}\ }\textbf {\bibinfo
  {volume} {101}},\ \bibinfo {pages} {043522} (\bibinfo {year} {2020})},\
  \Eprint {http://arxiv.org/abs/1910.12357} {arXiv:1910.12357 [hep-ph]}
  \BibitemShut {NoStop}%
\bibitem [{\citenamefont {Cutting}\ \emph {et~al.}(2018)\citenamefont
  {Cutting}, \citenamefont {Hindmarsh},\ and\ \citenamefont
  {Weir}}]{Cutting:2018tjt}%
  \BibitemOpen
  \bibfield  {author} {\bibinfo {author} {\bibfnamefont {D.}~\bibnamefont
  {Cutting}}, \bibinfo {author} {\bibfnamefont {M.}~\bibnamefont {Hindmarsh}},
  \ and\ \bibinfo {author} {\bibfnamefont {D.~J.}\ \bibnamefont {Weir}},\
  }\href {\doibase 10.1103/PhysRevD.97.123513} {\bibfield  {journal} {\bibinfo
  {journal} {Phys. Rev. D}\ }\textbf {\bibinfo {volume} {97}},\ \bibinfo
  {pages} {123513} (\bibinfo {year} {2018})},\ \Eprint
  {http://arxiv.org/abs/1802.05712} {arXiv:1802.05712 [astro-ph.CO]}
  \BibitemShut {NoStop}%
\bibitem [{\citenamefont {Seto}\ \emph {et~al.}(2001)\citenamefont {Seto},
  \citenamefont {Kawamura},\ and\ \citenamefont {Nakamura}}]{Seto:2001qf}%
  \BibitemOpen
  \bibfield  {author} {\bibinfo {author} {\bibfnamefont {N.}~\bibnamefont
  {Seto}}, \bibinfo {author} {\bibfnamefont {S.}~\bibnamefont {Kawamura}}, \
  and\ \bibinfo {author} {\bibfnamefont {T.}~\bibnamefont {Nakamura}},\ }\href
  {\doibase 10.1103/PhysRevLett.87.221103} {\bibfield  {journal} {\bibinfo
  {journal} {Phys. Rev. Lett.}\ }\textbf {\bibinfo {volume} {87}},\ \bibinfo
  {pages} {221103} (\bibinfo {year} {2001})},\ \Eprint
  {http://arxiv.org/abs/astro-ph/0108011} {arXiv:astro-ph/0108011} \BibitemShut
  {NoStop}%
\bibitem [{\citenamefont {Reichert}\ \emph {et~al.}(2022)\citenamefont
  {Reichert}, \citenamefont {Sannino}, \citenamefont {Wang},\ and\
  \citenamefont {Zhang}}]{Reichert:2021cvs}%
  \BibitemOpen
  \bibfield  {author} {\bibinfo {author} {\bibfnamefont {M.}~\bibnamefont
  {Reichert}}, \bibinfo {author} {\bibfnamefont {F.}~\bibnamefont {Sannino}},
  \bibinfo {author} {\bibfnamefont {Z.-W.}\ \bibnamefont {Wang}}, \ and\
  \bibinfo {author} {\bibfnamefont {C.}~\bibnamefont {Zhang}},\ }\href
  {\doibase 10.1007/JHEP01(2022)003} {\bibfield  {journal} {\bibinfo  {journal}
  {JHEP}\ }\textbf {\bibinfo {volume} {01}},\ \bibinfo {pages} {003} (\bibinfo
  {year} {2022})},\ \Eprint {http://arxiv.org/abs/2109.11552} {arXiv:2109.11552
  [hep-ph]} \BibitemShut {NoStop}%
\bibitem [{\citenamefont {Borsanyi}\ \emph {et~al.}(2016)\citenamefont
  {Borsanyi} \emph {et~al.}}]{Borsanyi:2016ksw}%
  \BibitemOpen
  \bibfield  {author} {\bibinfo {author} {\bibfnamefont {S.}~\bibnamefont
  {Borsanyi}} \emph {et~al.},\ }\href {\doibase 10.1038/nature20115} {\bibfield
   {journal} {\bibinfo  {journal} {Nature}\ }\textbf {\bibinfo {volume}
  {539}},\ \bibinfo {pages} {69} (\bibinfo {year} {2016})},\ \Eprint
  {http://arxiv.org/abs/1606.07494} {arXiv:1606.07494 [hep-lat]} \BibitemShut
  {NoStop}%
\end{thebibliography}%

\clearpage
\onecolumngrid
\begin{center}
   \textbf{\large SUPPLEMENTAL MATERIAL \\[.1cm] ``Signatures of a High Temperature QCD Transition in the Early Universe''}\\[.2cm]
  \vspace{0.05in}
  {Philip Lu, Volodymyr Takhistov, George M. Fuller}
\end{center}

\twocolumngrid
\setcounter{equation}{0}
\setcounter{figure}{0}
\setcounter{table}{0}
\setcounter{section}{0}
\setcounter{page}{1}
\makeatletter
\renewcommand{\theequation}{S\arabic{equation}}
\renewcommand{\thefigure}{S\arabic{figure}}
\renewcommand{\thetable}{S\arabic{table}}

\onecolumngrid

In this Supplemental Material we discuss details of PNJL model for a high temperature QCD transition and computation of parameters such as equation of state for scenarios
considered in the main text. 
 
\section*{PNJL Model for High Temperature QCD Transition}
 
In the following, we follow Refs.~\cite{Ratti:2005jh,Helmboldt:2019pan,Reichert:2021cvs} to construct effective PNJL model description of a high temperature QCD transition. 

The effective Lagrangian consists of three parts, the quark/meson sector, the gauge sector, and the interaction sector~\cite{Ratti:2005jh}:
\begin{equation}
\label{eq:efflagrange}
    \mathcal{L}_{PNJL} = \Bar{\chi}\left(i\gamma_\mu D^\mu - m_0\right)\chi + \frac{G}{2}\left[(\chi \Bar{\chi})^2+(\Bar{\chi}i\gamma_5 \vec{\tau} \chi)^2\right] - \mathcal{U}(\Phi,\Bar{\Phi},T)~,
\end{equation}
where $m_0=5.5\textrm{ MeV}$ is the current quark mass, $G=10.08\textrm{ GeV}^{-2}$ the coupling,  and $\mathcal{U}$ the Polyakov loop contribution.  The constituent quark mass is composed to the current quark mass and the meson field expectation value 
\begin{equation}
m=m_0-G\langle \Bar{\chi}\chi\rangle = m_0 - \sigma~.
\end{equation}
In the limit of zero chemical potential ($\Phi=\Bar{\Phi}$), the potential $\Omega$ as a function of the meson field $\sigma$ and Polyakov loop $\Phi$ is given by
\begin{align}
\begin{split}
\label{eq:effpotential}
    \Omega =&~ U(\Phi,T) + \frac{\sigma^2}{2G}  - 6 N_f \int_{0}^{\Lambda} \frac{d^3 p }{(2\pi)^3}E_p \\ &- 4 N_f T \int_{0}^{\infty} \frac{d^3 p}{(2\pi)^3}\ln\left[1+3\Phi e^{-E_p/T}+3\Phi e^{-2 E_p /T}+e^{-3E_p /T}\right]~,
\end{split}
\end{align}
where $N_f$ is the number of massless quark flavors. The pure gauge contribution is
\begin{align}
\begin{split}
\label{eq:gaugecont}
    & U(\Phi,T) = T^4\left[-\frac{\Phi^2}{2} \left(a_0 + a_1 \left(\frac{T_0}{T}\right) + a_2 \left(\frac{T_0}{T}\right)^2 + a_3 \left(\frac{T_0}{T}\right)^3\right) - \frac{b_3}{3}\Phi^3 + \frac{b_4}{4}\Phi^4\right]~,
\end{split}
\end{align}
where $a_i$ and $b_i$ are phenomenological coefficients. We employ polynomial fit to data of the Polyakov loop potential provided in Ref.~\cite{Ratti:2005jh} and consider $N_c=3$, resulting in $ a_0 = 6.75$, $a_1 = -1.95$, $a_2=2.625$, $a_3 = -7.44$, $b_3 = 0.75$, $b_4 = 7.5$. Here $T_0$ is the critical temperature of the confinement phase transition. 

The expectation values of $\sigma$ and $\Phi$ are the critical points of $\Omega$, and can be solved for by setting
\begin{align}
\begin{split}
\label{eq:sigmavar}
    \frac{d\Omega}{d\sigma} =&~ \frac{\sigma}{G} - 12N_f T \int \frac{d^3 p}{(2\pi)^3} \frac{\frac{\Phi m}{E_p T} e^{-E_p/T} + \frac{2\Phi m}{E_p T} e^{-2E_p/T} + \frac{m}{E_p T} e^{-3E_p/T}}{1+3\Phi e^{-Ep/T}+3\Phi e^{-2E_p/T}+e^{-3E_p/T}} \\ &+ \frac{3 N_f}{2\pi^2}\left(\Lambda\sqrt{\Lambda^2+m^2}-m^2 \tanh^{-1}\left[\frac{\Lambda}{\sqrt{\Lambda^2+m^2}}\right] \right) = 0~\\
    \frac{d\Omega}{d\Phi} =&~ T^4\left[-\Phi \left(a_0 + a_1 \left(\frac{T_0}{T}\right) + a_2 \left(\frac{T_0}{T}\right)^2 + a_3 \left(\frac{T_0}{T}\right)^3 + a_4 \left(\frac{T_0}{T}\right)^4\right) - b_3\Phi^2 + b_4\Phi^3\right] \\ 
     &- 12N_f T \int \frac{d^3 p}{(2\pi)^3} \frac{e^{-E_p/T}+e^{-2E_p/T}}{1+3\Phi e^{-E_p/T}+3\Phi e^{-2E_p/T}+e^{-3E_p/T}}=0~.
\end{split}
\end{align}
We numerically solve for $\sigma(T)$ and $\Phi(T)$ at each temperature for a given set of $T_0$, $\Lambda$, and $N_f$, which can then be used in Eq.~\eqref{eq:effpotential}.

\begin{figure*}[t]
\begin{center}
\includegraphics[width=0.45\columnwidth]{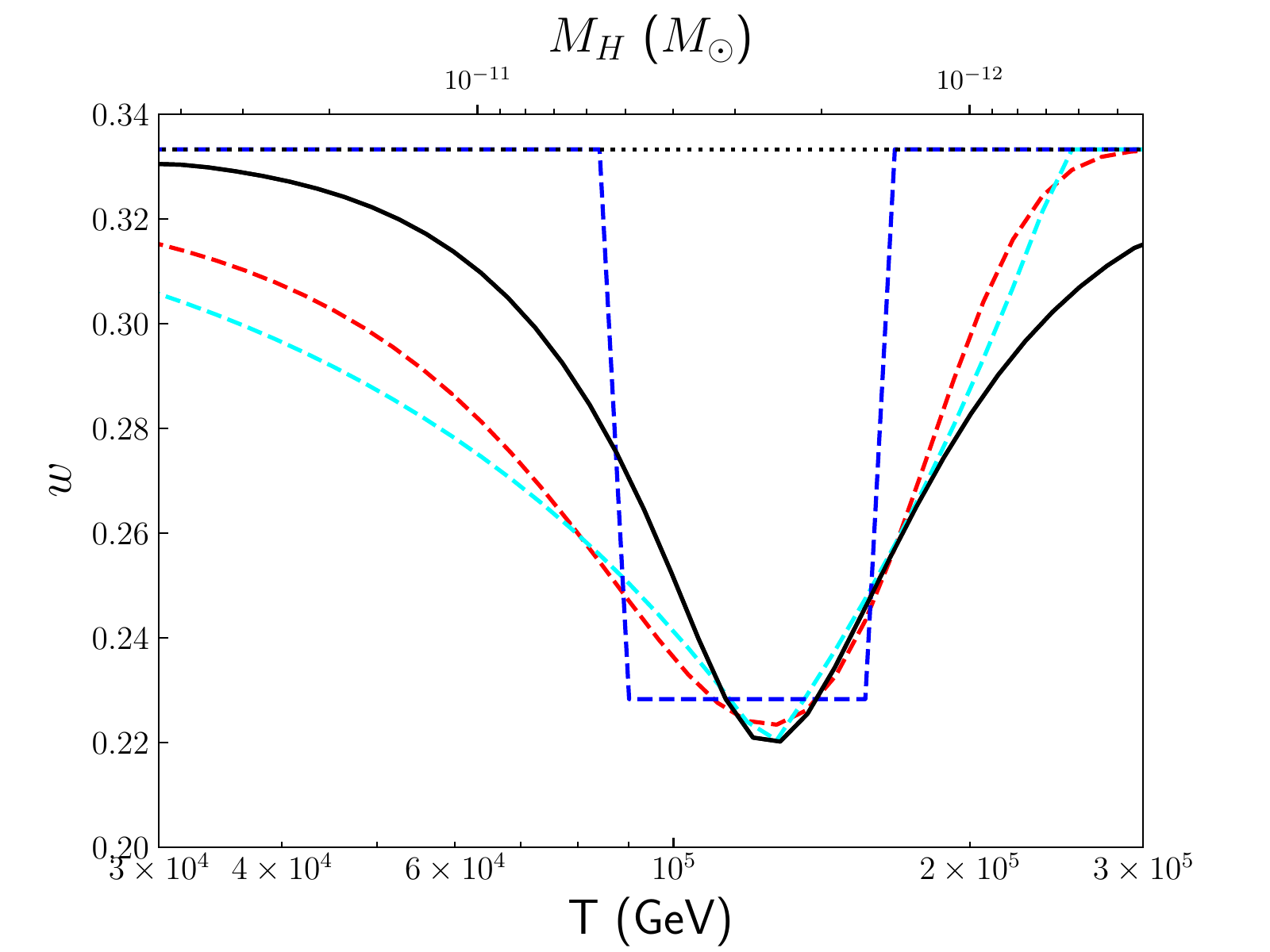}
\includegraphics[width=0.45\columnwidth]{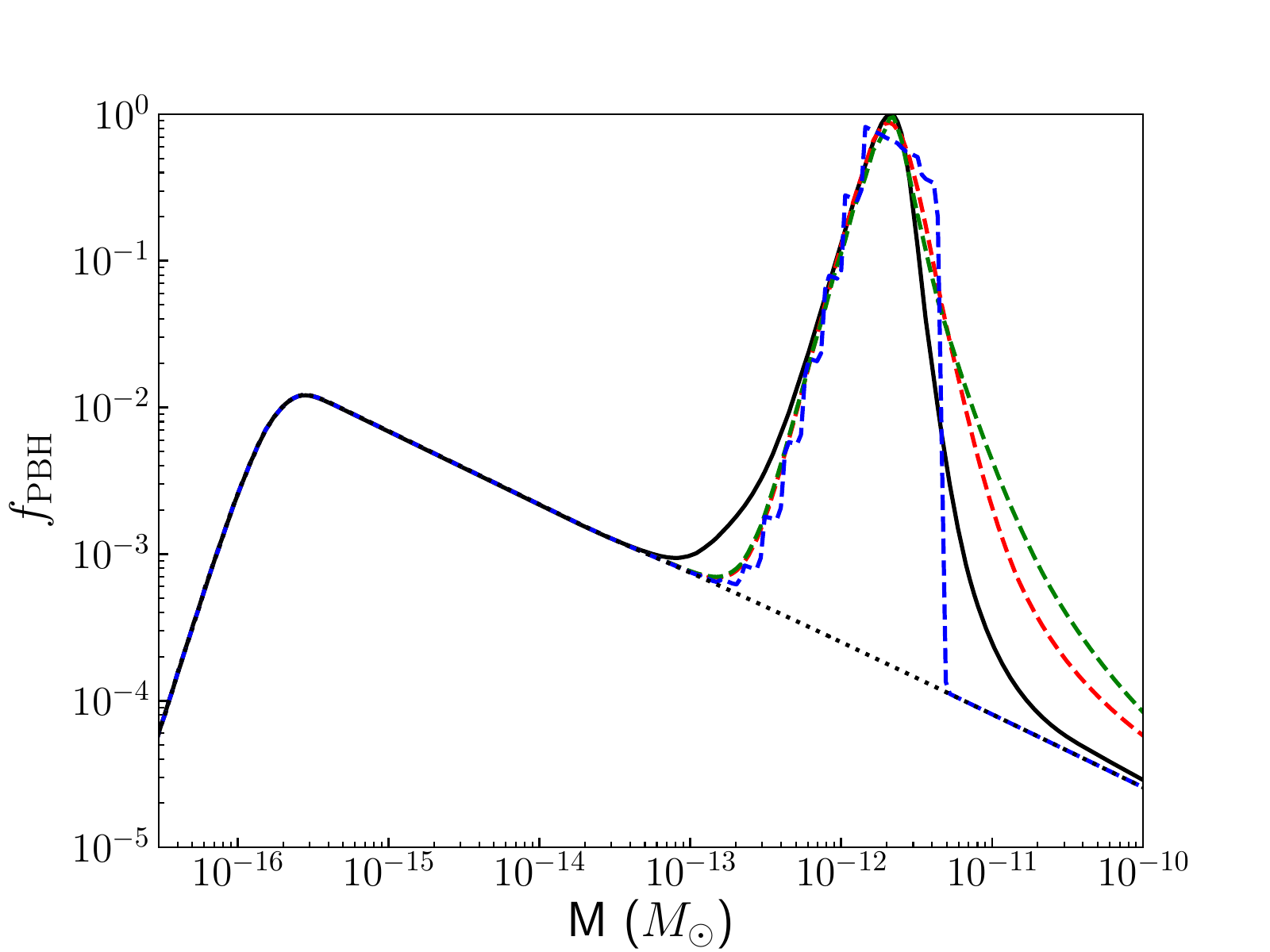}  
\caption{
[Left] Temperature evolution of EoS parameter $w$ due to a high temperature QCD transition with a critical transition temperature of $T_0 \sim 150$~TeV. [Right] We show PBHs contributing to DM from a first order QCD transition at high scales. 
Parameters of Model A of main text Table I are shown here with the PNJL (solid black), Gaussian (dashed red), Triangular (dashed green), Square (dashed blue) and Flat (dotted black) equations of state parameter $w$.
}
\label{fig:eosmodels}
\end{center} 
\end{figure*} 

Using thermodynamic relations~\cite{Ratti:2005jh}, the pressure and energy density are given by
\begin{align}
p =&~\dfrac{\partial F}{\partial V} =-\Omega \notag\\
\rho =&~T \dfrac{\partial^2F}{\partial T \partial V} - \dfrac{\partial F}{\partial V} = \Omega-T\dfrac{\partial \Omega}{\partial T}
\end{align}
 The equation of state parameter $w=p/\rho$ is evaluated as a function of temperature. We separately include the contributions from the non-QCD SM components (i.e. electrons, neutrinos, photons, etc.) that exert normal relativistic gas pressure $p=\rho/3$. The combined equation of state parameter is
\begin{equation}
    w = \frac{p + \rho_{SM-QCD}/3}{\rho + \rho_{SM-QCD}}~.
\end{equation}
We take $\rho_{SM-QCD} = (\pi^2 g_{SM-QCD}/30) T^4$ with $g_{SM-QCD} = 27.75$ at high temperatures $T>200$ GeV. We have estimated $g_{SM-QCD}$ above $30$~MeV by subtracting out the relevant QCD degrees of freedom from the data of Ref.~\cite{Borsanyi:2016ksw}.

In Fig.~\ref{fig:eosmodels} we display the evolution of EoS parameter $w$ as a function of T and horizon mass $M_H$ for PNJL parameters allowing for PBHs in the open asteroid-mass range. Further, we include several additional consider phenomenological parametrizations of $w$ behavior for reference:  
\begin{align*}
\begin{split}
\label{eq:wmodels}
    Flat:~~~w~&=\frac{1}{3} \\
    Gaussian:~~~w~&=\frac{1}{3}-0.11\exp{-\frac{T-125\textrm{ GeV})^2}{2 (50\textrm{ GeV})^2}} \\
    Triangular:~~~w~&=\frac{1}{3}-(0.115-0.115\frac{T-125\textrm{ GeV}}{125\textrm{ GeV}})\Theta\left((0.115-0.115\frac{T-125\textrm{ GeV}}{125\textrm{ GeV}}\right) \\
    Square:~~~w~&=\frac{1}{3}-0.105\Theta\left(T-85\textrm{ TeV}\right)\Theta\left(165\textrm{ GeV}-T\right)
\end{split}
\end{align*}

\begin{figure*}[t]
\begin{center}
\includegraphics[width=0.45\columnwidth]{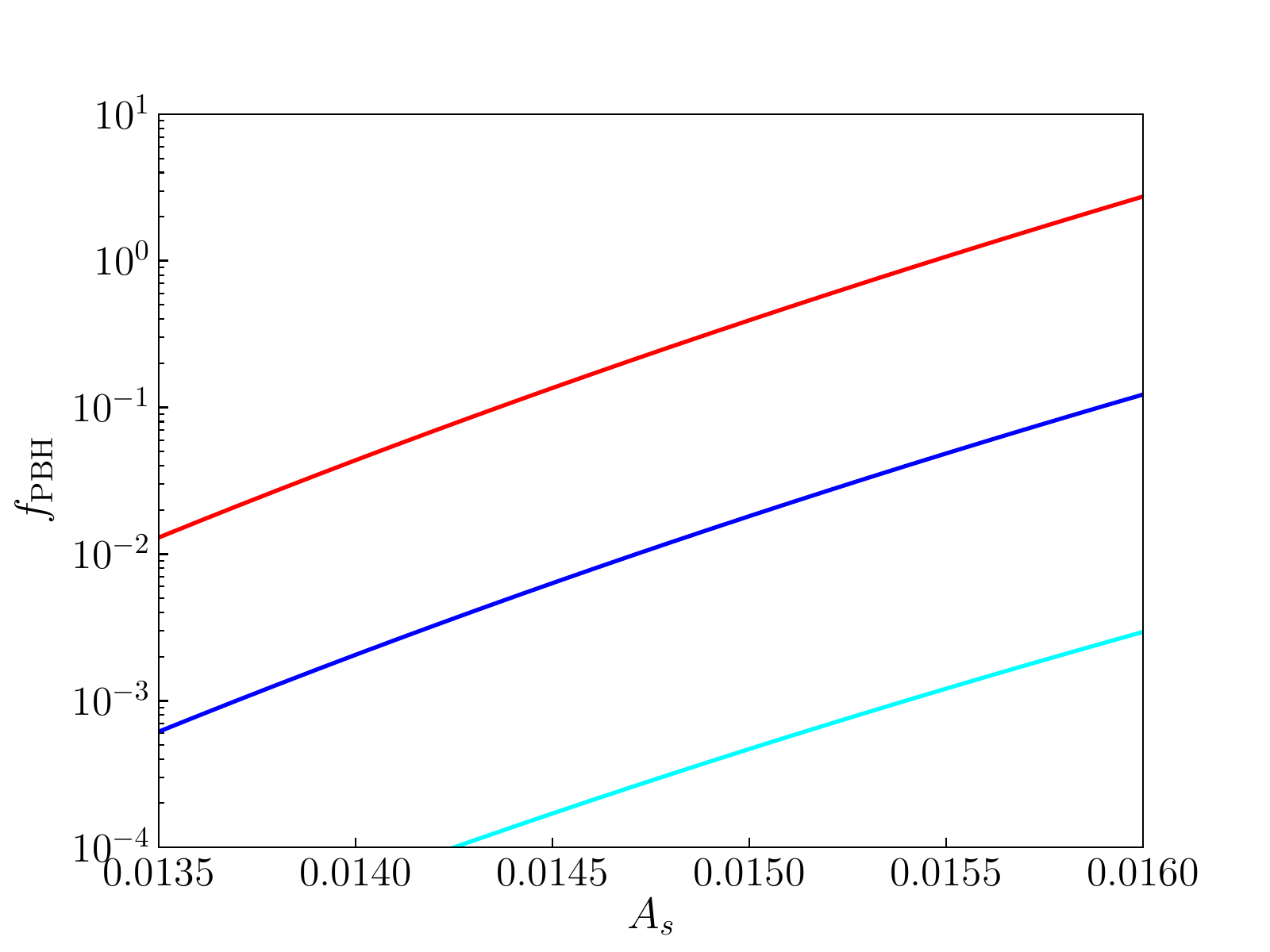}
\caption{
Variation of $f_{\rm PBH}$ as a function of the primordial power spectrum amplitude $A_s$ that results in PBHs, for Models A (red), B (blue) and C (cyan) discussed in the main text Table I.
}
\label{fig:fbhas}
\end{center} 
\end{figure*}

In Fig.~\ref{fig:fbhas} we display how $f_{\rm PBH}$ varies with primordial power spectrum amplitude $A_s$ for Models A, B and C of the main text Table I.

\end{document}